\documentclass[preprint,12pt]{elsarticle}
\usepackage{amssymb}
\usepackage{amsmath}
\usepackage{amsthm}
\usepackage{float}
\usepackage{comment}
\usepackage{titlecaps}
\usepackage[table,xcdraw]{xcolor}
\usepackage{tikz}
\usetikzlibrary{arrows,shadows} 
\usetikzlibrary{plotmarks}
\usetikzlibrary{shapes,shadows,arrows,fit}
\usetikzlibrary{decorations.pathmorphing}
\usetikzlibrary{decorations.markings}
\usetikzlibrary{calc} 
\usetikzlibrary{positioning,chains,scopes}
\usetikzlibrary{shapes.geometric,shapes.arrows,decorations.pathmorphing}
\usetikzlibrary{matrix,chains,scopes,positioning,arrows,fit}

\usepackage{algorithmic}
\usepackage{nicefrac}
\usepackage{mathtools}
\usepackage[linesnumbered,lined,boxed,commentsnumbered,ruled]{algorithm2e}
\usepackage[mathscr]{eucal}
\usepackage{booktabs}
\usepackage{siunitx}
\usepackage{array, multirow}
\usepackage{nicefrac}
\usepackage{subcaption} 
\usepackage{natbib}
\RequirePackage[pdftex,
      colorlinks={true},
      linkcolor={Blue3},
      urlcolor={Blue3},
      citecolor={Blue3},
      pdfstartview={FitH},
      bookmarks={true},
            pdfauthor={Author},
            pdftitle={Title},
            pdfsubject={Subject}
            plainpages=false,
            pdfpagelabels
            ]{hyperref}

\newenvironment{RQquestion}{%
  \par%
  \leftskip=1em\rightskip=1em%
  \noindent
  \ignorespaces}
  
\newenvironment{RQanswer}{%
  \par%
  \leftskip=3em\rightskip=2em%
  \noindent
  \ignorespaces}{%
  \par\medskip}
  
\journal{Journal of Systems \& Software}

\begin{document}
\let\WriteBookmarks\relax
\def\floatpagepagefraction{1}
\def\textpagefraction{.001}

\title{Wayback Machine: \textcolor{black}{A tool to capture} the evolutionary behaviour of \textcolor{black}{the bug reports and their triage process} in open-source software systems}

\author[1]{Hadi Jahanshahi\corref{cor1}}
\ead{hadi.jahanshahi@ryerson.ca}
\author[1]{Mucahit Cevik}
\author[2]{José Navas-Sú}
\author[1]{Ay\c{s}e Ba\c{s}ar}
\author[2]{Antonio González-Torres}


\address[1]{Data Science Lab at Ryerson University, Toronto, Canada}
\address[2]{Computer Engineering Dep. at Costa Rica Institute of Technology, Cartago, Costa Rica}
\cortext[cor1]{Corresponding author}

\begin{abstract}
The issue tracking system (ITS) is a rich data source for data-driven decision-making. Different characteristics of bugs, such as severity, priority, and time to fix, provide a clear picture of an ITS. Nevertheless, such information may be misleading. For example, the exact time and the effort spent on a bug might be significantly different from the actual reporting time and the fixing time. Similarly, these values may be subjective, e.g., severity and priority values are assigned based on the intuition of a user or a developer rather than a structured and well-defined procedure. \textcolor{black}{Hence, we explore the evolution of the bug dependency graph together with priority and severity levels to explore the actual triage process. Inspired by the idea of the ``Wayback Machine'' for the World Wide Web, we aim to reconstruct the historical decisions made in the ITS. Therefore, any bug prioritization or bug triage algorithms/scenarios can be applied in the same environment using our proposed ITS Wayback Machine. More importantly, we track the evolutionary metrics in the ITS when a custom triage/prioritization strategy is employed.} 
\textcolor{black}{We test the efficiency of the proposed algorithm using data extracted from three open-source projects. Our empirical study sheds light on the overlooked evolutionary metrics~\textendash~e.g., overdue bugs and developers' loads~\textendash~which are facilitated via our proposed past-event re-generator.}

\end{abstract}

\begin{highlights}
\item Issue tracking systems are dynamic and confronted with the uncertainties of bug reports.
\item We propose a Wayback Machine to explore past bug prioritization and \textcolor{black}{triage} approaches.
\item The Wayback Machine enables practitioners \textcolor{black}{to investigate the evolution of the bug dependency graph and bug features.}
\item \textcolor{black}{Researchers can evaluate their bug triage or prioritization methods using the Wayback Machine.} 
\item \textcolor{black}{It provides a complete list of traditional and evolutionary metrics for a given triage or prioritization method.}
\end{highlights}


\maketitle


\section{Introduction}
In software engineering practice, the later the bug is discovered and fixed, the more costly it will be for the projects~\cite{kumar2017software}. However, due to limited resources and an increasing number of defects reported, it is typically not feasible to fix all the bugs before each software release. Therefore, practitioners frequently face the decision of which bugs to resolve now or defer to the next release. 

Bug prioritization \textcolor{black}{and triage} tasks mainly depend on the quality of the reported bugs while noting that not all reports are consistent. Previous studies discussed the evidence for the mismatch between developers' and users' understanding of the bugs~\cite{Bettenburg2008, umer2019}. 
\textcolor{black}{Moreover, as expected, bug severity information is typically not deemed to be reliable as evidenced by 51\% of the duplicate reported bugs having inconsistent severity labels~\cite{tian2016}.}
Data on the bug fixing time is not reliable either; that is, it does not indicate the exact amount of working hours on a specific bug in a continuous manner~\cite{Shirin2020}. 

Unlike many subjective characteristics of the bugs, blocking bugs are determined by a developer in the phase of defect resolution. In a typical flow of a bug report, an end-user or a developer reports a bug or an issue. Subsequently, a triager assigns it to a developer, or a developer claims its possession. Ultimately, after they find a resolution for the bug, it is verified by another developer and gets closed. However, in the case of a blocking bug, the process is interrupted~\cite{Garcia2018}. Blocking bugs have higher complexity than non-blocking bugs, require more time to get fixed, associate with a larger codebase (e.g., in terms of lines of code), and are also difficult to be predicted~\cite{Garcia2018, Goyal2017, Wang2018}. As the number of blocking bugs increases, resource planning becomes a tedious task, and developers defer many bugs to a later release. Accumulation of lingering bugs~\textendash~the bugs reported but not resolved in the current release~\textendash~degrades the software quality and increases the maintenance cost~\cite{Akbarinasaji2017}. Therefore, understanding the influence of the bug dependency graph (BDG) \textcolor{black}{together with other bug features} is essential for software maintenance.

\textcolor{black}{A common approach to bug triage and prioritization is to use different machine learning algorithms and find the performance of the bug assignment or bug prioritization~\cite{uddin2017, alenezi2013efficient,anvik2006should,xuan2017automatic}. However, in most cases, previous studies did not consider the effect of bug dependency in their recommended solution.
Moreover, for a fair performance assessment, it is important to explore the impact of the algorithm at the exact time a bug is assigned or prioritized. 
For instance, at time $t$, if a bug is assigned to a developer having previous experience with the associated component but busy with other assigned tasks, the algorithm is expected to automatically propose an alternative developer for the open bug. 
However, without a simulator that regenerates the exact characteristics of the open bugs and available developers at time $t$, it might not be feasible to propose a practical solution. 
Accordingly, we propound the modular Wayback Machine that regenerates past events for any given timestamp and might be easily adopted by researchers to investigate the performance of their proposed bug prioritization or triage algorithm. }

Another important missing link in the previous studies is to recognize the actual situation in the real world as a baseline. 
It is critical to know how the \textcolor{black}{content of} the issue tracking system (ITS) evolves in terms of complexity as it enables practitioners to automate the decision-making process and to trace back the actual decisions in the bug triage process. \textcolor{black}{The idea of the Wayback Machine comes from the digital archive of the World Wide Web, which enables exploring the status and content of the webpages in previous timestamps\footnote{\hyperlink{https://archive.org/web/}{https://archive.org/web/}}. To this end, we construct a Wayback Machine with which practitioners may explore the past events in the ITS.} Besides, we examine an extensive list of \textcolor{black}{prioritization and triage strategies over a BDG to explore whether the proposed event-regenerator machine can reveal evolutionary aspects of decisions that were not explored in previous studies.} 
Additionally, we consider using a discrete-event system simulation approach and evaluate the performance of the models using \textcolor{black}{both traditional metrics (e.g., the assignment accuracy) and evolutionary (e.g., the task concentration on developers)} metrics. 
Accordingly, our research questions are two-fold: first, understanding and rebuilding the history of the ITS; and second, \textcolor{black}{checking the validity of the Wayback Machine through exploring prioritization and triage strategies. 
We note that these strategies can be substituted with any bug prioritization or triage algorithm in the modular Wayback Machine. Thus}, we structure our study along with the following five research questions, divided into two categories:

\begin{RQquestion}
    \textbf{RQ1a: How do open-source software systems evolve in terms of the number of bug reports, bug dependencies, and lingering bugs?}
\end{RQquestion}
\begin{RQanswer}
    We explore the past events in the ITS through a novel Wayback Machine. Given the extracted data from any ITS, this machine provides us with bugs' status in any timestamp. Hence, we may query different characteristics of the bugs and explore the reason behind each bug prioritization decision in the past, e.g., what kinds of bugs existed and why a developer chose to resolve a specific bug over others. We demonstrate the number of bug reports, the evolution of BDG, and their effect on the lingering bugs. 
\end{RQanswer}

\begin{RQquestion}
    \textbf{RQ1b: How do the characteristics of the resolved bugs change over time?}
\end{RQquestion}
\begin{RQanswer}
    We further explore the importance of bug dependencies for triagers. We analyze a series of observed sequences through the Wayback Machine to see how triagers regard a bug's \textcolor{black}{severity, priority,} degree, and depth when prioritizing it. Our findings illustrate \textcolor{black}{that in some issue tracking systems, the dependency of the bugs is mainly disregarded, or even some developers are not aware of it, and so dependency loses its importance. On the other hand, in the ITS where bug dependency practice is taken seriously, the principal role of depth and degree is noticeable by comparing their average for both solved and postponed bugs. We also observe that although severity and priority levels are known to be subjective, the average severity and priority of the fixed bugs are higher than the open bugs in the ITS.} 
\end{RQanswer}

\begin{RQquestion}
    \textbf{\textcolor{black}{RQ2a: How do different bug prioritization strategies perform in terms of evolutionary metrics?}}
\end{RQquestion}
\begin{RQanswer}
    After creating the Wayback Machine to review past prioritization decisions, we explore different prioritization strategies and compare their performance with the actual case. \textcolor{black}{The main aim of RQ2s is to validate the proposed Wayback Machine as a way to prioritize bugs via different machine learning and rule-based approaches. To this end, we first define evolutionary metrics for the first time (e.g., the depth and degree of the BDG and the deviation from the actual assignment). We cannot report these metrics using machine learning algorithms in a traditional way due to the static nature of training and testing \textendash i.e., training a model on tabular information and reporting the performance without time consideration. As such, we evaluate different rule-based and machine learning algorithms for bug prioritization \textcolor{black}{purposes}.} 
\end{RQanswer}

\begin{RQquestion}
    \textbf{\textcolor{black}{RQ2b: How do different bug triage strategies perform in terms of evolutionary metrics?}}
\end{RQquestion}
\begin{RQanswer}
    \textcolor{black}{We further explore the performance of well-established bug triage algorithms. We equip the Wayback Machine with a bug triage module, which can compare existing triage algorithms in the literature with the actual bug assignment. Moreover, we report the performance of those algorithms based on the evolutionary and traditional metrics, i.e., static accuracy-related metrics.} 
\end{RQanswer}

We organized the rest of the paper as follows.  \textcolor{black}{Section~\ref{sec:research-methodology} presents the methodology, motivating example, and dataset description. Section~\ref{sec:Wayback} briefly explains the notion behind the Wayback Machine, and Section~\ref{sec:BugPrioritizationTriage} summarizes bug prioritization and triage problems.
Section~\ref{sec:findings} investigates the impact of different prioritization and triage strategies that take into account the evolutionary characteristics of the ITS. It reports the performance of the models based on both traditional and evolutionary metrics. Finally, Section~\ref{sec:threats} describes the limitations and threats to validity, followed by Section~\ref{sec:background}, which briefly discusses the relevant literature on bug prioritization, triage, and dependency graphs, and Section~\ref{sec:conclusion}, which concludes the paper.}

\section{Research methodology}\label{sec:research-methodology}
We examine the evolution of the bugs in the software repositories to contribute to understanding the bug prioritization and triage process. For this purpose, we use reported bug information extracted from the ITS of three open-source projects, namely, Mozilla, Eclipse, and LibreOffice, covering ten years (from January 2010 to December 2019). We construct a BDG based on the daily reported bugs (nodes) and daily blocking information (arcs). A BDG is a directed acyclic graph that does not contain any loop in terms of blocking information, i.e., a bug cannot block another bug and be blocked by the same bug simultaneously.

We track the BDG's evolution through various complexity metrics such as \textit{depth} ($\theta$) of a node that is the longest directed path between the given node and other nodes in the graph, the \textit{degree} ($\delta$) of a node that corresponds to the number of its outgoing arcs, the number of nodes ($n$), and the number of arcs ($m$) in a graph. Accordingly, the maximum degree and depth of a graph cannot exceed $n-1$. As we sort all the information chronologically, we start adding or removing nodes and arcs at each timestamp and measuring the changes in the metrics from time $t$ to time $t+1$. The accumulated information uncovers the evolution of the BDG in the project. More details about the BDG are provided in Section~\ref{sec:motivating_example}.

To accurately trace back the history of the actual software project, we also incorporate bug report attributes such as bug title, description, severity, and priority. \textcolor{black}{We further utilize these attributes and create machine learning algorithms and rule-based approaches to validate the Wayback Machine in a controlled experiment.} Also, we regenerate the network's behavior using different bug prioritization and triage strategies and compare them in terms of various \textcolor{black}{traditional and evolutionary metrics}. 


\subsection{\textcolor{black}{Motivating example}}\label{sec:motivating_example}

\textcolor{black}{The Wayback Machine makes it possible to evaluate/observe the evolution of a project as it records the events in the ITS and generates evolutionary statistics such as the number of reported/fixed bugs, their relevant severity, priority, depth and degree, together with information on the developers' load. \textcolor{black}{These evolutionary statistics} are time-reliant and may reveal changes from one release to another. \textcolor{black}{Accordingly, we list three essential aspects of bug prioritization and triage decisions that are not considered \textcolor{black}{simultaneously} in the previous studies: bug dependency, time, and decision outcome. The current work incorporates all three aspects simultaneously.} Here, we discuss the importance of covering each of them in bug/defect prioritization/triage studies. We note that the Wayback Machine covers those aspects by design.}

\begin{figure}[!ht]
\centering
    \begin{tikzpicture}[b/.style={circle,draw,execute at begin node={$b_{#1}$},
        alias=b-#1,label={[rectangle,draw=none,overlay,alias=l-#1]right:{$[s_{#1},c_{#1}]$}}}]
    \node[matrix of nodes,column sep=1em,row sep=2em]{
     & & |[b=1]|& & |[b=2]| & & &|[b=7]|\\
     &  |[b=3]|& & |[b=4]| &  & & & |[b=8]| \\
       |[b=5]|& & |[b=6]| &  & & &|[b=9]| &\\
    };
    \path[-stealth] foreach \X/\Y in {1/3,3/6,1/4,4/6,2/4,7/8} {(b-\X) edge (b-\Y)};
    \path (l-7.east); 
    \end{tikzpicture}
    \caption{\textcolor{black}{A typical BDG, with severity ($s_i$) and fixing time ($c_i$) for each bug $b_i$.}}
    \label{fig:BDG}
\end{figure}
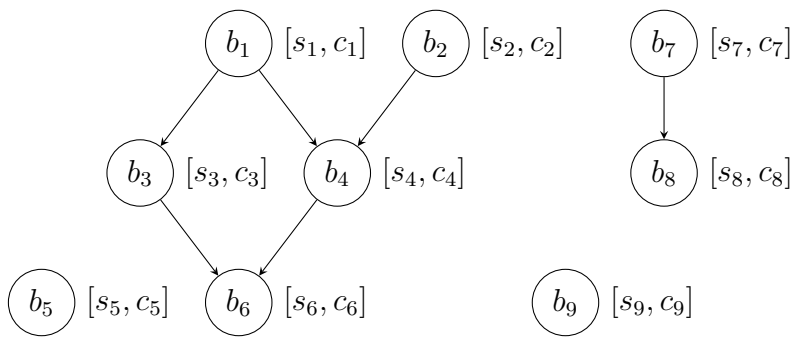

\paragraph{Bug dependency}
\textcolor{black}{Figure~\ref{fig:BDG} shows the dependency graph of the bugs, $b_i \in \{b_1, b_2, \dots, b_9\}$ with their associated severity, $s_i$, and the fixing time, $c_i$. 
Nodes show the bugs, and arcs show their dependencies determined by developers. 
\textcolor{black}{We consider the parent nodes to be the blocking bugs and child nodes to be the blocked bugs, and the direction of the arcs in a BDG determines the relationship between the bugs.
For instance, $b_1$ is the parent of $b_3$ and $b_4$, and $b_2$ is the parent of $b_4$.
}Since $b_1$ and $b_2$ are blocking bugs for $b_4$ (i.e., child node), it cannot be solved unless its blocking bugs \textcolor{black}{(i.e., parent nodes) are fixed. 
}In a sparse BDG, we may observe a plethora of solo bugs (e.g., see $b_5$ and $b_9$), which neither block nor are blocked by others. On the other hand, having many blocked bugs in the system may postpone the bug fixing process and impose lingering bugs in the system~\cite{Shirin2020}. If triagers disregard the dependency of the bugs while prioritizing them, they may arrive at a decision that is infeasible in practice that might cause delays in bug resolution times.} 

\textcolor{black}{Other important factors in a BDG include the number of subgraphs and depth and degree values for the bugs. 
In this paper, we simply refer to out-degree as \textit{degree}. Figure~\ref{fig:BDG} has four subgraphs, $\mathcal{S} = \{[1,2,3,4,6],[5],[7,8],[9]\}$. Also, $b_6$ has the highest depth value of 2, and $b_1$ has the highest degree value of 2. A degree shows the number of blocked bugs, and depth indicates the number of parents and grandparents of a bug in a graph. A higher depth of a bug may lead to its fixing time postponement due to its many parents. Accordingly, we closely track the dependency of the bugs during the bug triage process.}
The historical data of Bugzilla for Mozilla, Eclipse JDT, and LibreOffice projects indicates many solo bugs, whereas, in the same projects, some densely connected sub-graphs gradually accumulate. Our evolutionary model, Wayback Machine, can trace back to when each of these sub-graphs developed. It provides a clear insight into the exact time when \textcolor{black}{an inappropriate prioritization/triage} resulted in either lingering bugs or an unbalanced network.

We note that, while dependency information is available in the software repositories (e.g., Bugzilla), only a few other studies considered dependency as an important factor while designing bug prioritization and triage algorithms. Accordingly, our study also contributes to a better understanding of the dependency information in bug prioritization and triage. \textcolor{black}{Moreover, bugs may have a mutual interaction, e.g., two bugs might need to be resolved simultaneously. However, we use the ``blocking'' and ``blocked'' information of the bugs from the repositories in which the mutual relationship is not clearly defined. Hence, we employ the available information without considering the requirement for simultaneous bug resolution.
In our framework, the bugs that need to be resolved jointly can be represented as a single bug in the database, ensuring simultaneous bug resolution.}

\paragraph{Time} \textcolor{black}{Another major factor in bug triage is time. Most studies on bug prioritization and triage that use bug history without simulation do not consider the evolutionary nature of the ITS~\citep{uddin2017, zaidi2020applying, alazzam2020automatic, park2011costriage}. For instance, if a model recommends solving bug $i$ prior to bug $j$ at time $t$, this recommendation should be made while other bugs and the information of the bug $i$ and $j$ are consistent with time $t$. The severity of bug $i$, $s_i$, changes over time. Therefore, if we consider an approach to use severity as a feature that may affect the bug prioritization, this severity should be the exact severity of the bug at time $t$. Moreover, the bug might not be blocked by another bug at time $t$, but it becomes blocked in future time steps. That is, we need to consider the exact dependency at the time of solving the bug. This logic can be generalized to any other evolutionary feature of a bug. Lastly, when prioritizing a bug, it is important to know the exact list of open bugs at that time.}

\paragraph{Decision outcome} \textcolor{black}{We cannot prioritize or triage all the available bugs without considering the opening, closing, and re-opening status. That is, only having high accuracy in bug assignment or prioritization does not guarantee that a model can be applied for the real world. For instance, assume that we assign bug $b_i$ to developer $d_j$ at time $t$. This assignment may be considered accurate as the developer has previous experience with bugs of the same type/component. However, the developer might be overloaded by previously assigned bugs and cannot claim possession of a new bug at time $t$. In such a case, a second developer who is fairly knowledgeable in the field can start working on the new bug to avoid bug accumulation in the ITS. Therefore, knowing the schedule and current loads of the developers might be very important. Accordingly, we define a set of evolutionary metrics such as the number of overdue bugs, which capture the real impact of a decision at each timestamp. We also check the assignment time of the developers and compare each strategy with the actual case to see whether the strategy mimics the real world. We note that all bug prioritization and triage algorithms in the literature may benefit from a stable past-event re-generator that captures the evolutionary history of the bugs. The ITS Wayback Machine, coded in Python, serves this purpose by its modular structure. Different bug prioritization or triage algorithms can be integrated into it, while the machine uses the chronological data and produces the visual and tabular outputs, giving more comprehensive insights into the decision outcomes. }

\textcolor{black}{\paragraph{Contribution} The Wayback Machine is a tool to reproduce past events in the ITS with their exact timestamp and track all the changes in the system. It does not involve any stochasticity/uncertainty (e.g., via random number generation) in its processes; instead, it employs the information extracted from the ITS, and runs them in the correct order. Overall, this tool contributes to the literature based on the following features and functionalities.
    \begin{itemize}
        \item The Wayback Machine regenerates past events (e.g., bug introduction, assignment, and resolution) using the data extracted from the ITS. These are the actual events/decisions made in the ITS without fitting any probability distribution to the event data. Accordingly, a user can retrieve the actual events in the system at any timestamp while using the Wayback Machine.
        \item The Wayback Machine provides a complete list of performance metrics and information of the changes in the ITS after making those decisions (see Section 4.2 for more details). For instance, it captures the dependency information of the bugs and constructs the BDG after each event occurs in the system. Accordingly, it follows the decision outcomes (e.g., changes in the depth and degree of the BDG or changes in the schedules and experiences of developers) when a bug is assigned to a developer.
        \item The Wayback Machine incorporates different well-established bug triage and prioritization algorithms to compare their performance with the actual prioritization/triage decisions (see Sections 4.3 and 4.4 for more details). Any researcher can employ the tool to add their proposed triage/prioritization algorithm and track the system status after their model assigns/prioritizes a bug. Then, they may compare their model results with those of the previous studies and the actual decisions.
    \end{itemize}
    The Wayback Machine incorporates time, dependency, and decision outcomes considerations simultaneously by carefully implementing the prioritization/triage decisions, and keeping track of the changes in the system.  
}

\textcolor{black}{\subsection{Current bug prioritization and triage practice in Bugzilla projects}
A newly reported bug to the Bugzilla ITS has an ``UNCONFIRMED'' status until it is validated. A developer starts ``preparation'' steps, i.e., searching for bugs according to their expertise, checking their information and Metadata, and finding possible duplicate bugs. After passing that phase, they try to reproduce the bug. If they confirm a bug based on its reproducibility, its status changes to ``NEW'' and becomes ready for the prioritization and assignment phase. \textcolor{black}{In some projects, bug triage meetings help developers review open bugs, evaluate whether each bug is worth fixing, decide when they should be fixed, and determine who should work on it.
On the other hand, this is not a common practice for some smaller open source projects (e.g., LibreOffice).
}Note that although the prioritization might be subjective, the QA team members need to be consistent in determining bugs prioritization and have a clear flowchart to set the priority level.
They might also flag a bug as ``UNCONFIRMED'', ``NEEDINFO'', and ``INVALID'' if a defect runs short of information or they fail to verify it. In Open Source Software (OSS) systems, in the case of critical bugs, the bug assignment is done by highlighting the bugs and CCing potential developers. As a result, a developer may claim possession of a verified bug rather than formally being assigned to it. Nevertheless, the practice of assigning a bug to a developer by a triager is another way of triaging bugs in the OSS\footnote{See triage for Bugzilla in \href{https://firefox-source-docs.mozilla.org/bug-mgmt/policies/triage-bugzilla.html}{Mozilla}, \href{https://wiki.documentfoundation.org/QA/BugTriage}{LibreOffice}, and \href{https://wiki.eclipse.org/SWT/Devel/Triage}{Eclipse} projects}.}

\subsection{Data collection}
We use bug data information from Bugzilla, an ITS for open-source software projects. The dataset is originally extracted from Mozilla, Eclipse, and LibreOffice ITSs and contains reported bugs for the projects between January 2010 and December 2019. We note that LibreOffice was forked in 2010 from OpenOffice, and its first reported bug was in August 2010. \textcolor{black}{According to the Bugzilla website\footnote{\hyperlink{https://www.bugzilla.org/installation-list}{https://www.bugzilla.org/installation-list}}, these projects are amongst the top-8 highlighted ``Free Software Projects'' and have a clear explanation of how to extract information from their repositories using API. There are many other projects to be considered, e.g., Linux Distribution projects; however, we choose these ones since they are diverse and well-established in terms of graph complexity, different number of reported bugs, and number of developers.} To collect the raw data from the repository, we use the Bugzilla REST API to extract both general information of all bugs and the history of all metadata changes for each bug\footnote{\hyperlink{https://wiki.mozilla.org/Bugzilla:REST_API}{https://wiki.mozilla.org/Bugzilla:REST\_API}}. The collected information includes the creation time, current status, type, severity, priority, title and description, resolution, assignee, and component. On the other hand, the evolutionary information is not obtainable via the general information of a bug. Consequently, we extract the formal relationship between the bugs by considering the metadata of their change history, along with their timestamps. These relationships take the form of duplication and blocking.

We examine both blocking and blocked bugs to see whether their initiation was before or after 2010. If a blocking or dependent bug was created before that time, we again extract all its information and add the ``old'' bug to the current database since they could affect the time to solve the corresponding bugs. Therefore, our database captures a full picture of bug dependency, whether it belongs to the targeted dates or earlier. For older bugs, we ignore the blocking information among themselves; however, we consider their dependency effects on targeted bugs between 2010 and 2020.

We next construct an evolutionary database. This database includes any change in the reported bugs along with their timestamps. Typically, these data cannot be obtained merely from bugs' information, and it requires extracting bugs' history as well. While extracting historical data from Bugzilla, we obtain both missing and contradictory information. We handle this issue by combining the information of duplicate bugs and their historical metadata changes. Lastly, we sort the events' logs by their timestamps and design a database that includes bugs' information in chronological order.

\subsection{Descriptive analysis}\label{sec:descriptive_analysis}
Table~\ref{tab:bug_info} shows the most relevant information regarding the extracted datasets. The number of publicly available bugs reported to Bugzilla between 2010 and 2020 for Mozilla, Eclipse, and LibreOffice is 100,475, 16,228, and 70,168, respectively. \textcolor{black}{These projects are diverse in terms of the number of reported bugs, the number of bug dependencies, and the ratio of open bugs to total reported bugs, hence deemed suitable for our analysis}. After extracting those bugs, we encounter some older bugs that block or are blocked by target bugs. We extract the information of the bugs older than 2010 if they are related to the target bugs. Therefore, our database includes the targeted bugs between 2010 and 2020 and older bugs before 2010. A complete report of their priority, severity, number of comments, and blocking information is provided in the table as well.

\begin{table}[!ht]
\centering
\caption{Information related to the bugs extracted from Bugzilla for Mozilla, Eclipse, and LibreOffice projects\label{tab:bug_info}}%
\resizebox{\linewidth}{!}{
\begin{tabular}{lrrrrr}
\toprule 
 \multicolumn{1}{c}{} & \multicolumn{2}{c}{\textbf{Mozilla}} & \multicolumn{2}{c}{\textbf{Eclipse}} & \multicolumn{1}{c}{\textbf{LibreOffice}} \\
 \cmidrule(lr){2-3}\cmidrule(lr){4-5}\cmidrule(lr){6-6}
 \textbf{Bug information} &  \multicolumn{1}{c}{\begin{tabular}[c]{@{}c@{}}\textbf{01/01/2010 -}\\ \textbf{31/11/2019}\\ \textbf{Targeted bugs}\end{tabular}} & \multicolumn{1}{c}{\begin{tabular}[c]{@{}c@{}}\textbf{09/06/1999 -} \\ \textbf{31/11/2009}\\ \textbf{Older bugs}\end{tabular}} & \multicolumn{1}{c}{\begin{tabular}[c]{@{}c@{}}\textbf{01/01/2010 -}\\ \textbf{31/11/2019}\\ \textbf{Targeted bugs}\end{tabular}} & \multicolumn{1}{c}{\begin{tabular}[c]{@{}c@{}}\textbf{09/06/1999 -} \\ \textbf{31/11/2009}\\ \textbf{Older bugs}\end{tabular}} & \multicolumn{1}{c}{\begin{tabular}[c]{@{}c@{}}\textbf{03/08/2010 -} \\ \textbf{31/11/2019}\\ \textbf{All bugs}\end{tabular}} \\
\midrule
\# of bugs & 100,475 & 12,944 & 
           16,228 & 114 & 
           70,168 \\
\midrule
Dependency info &&&&& \\
\quad \# of blocked bugs & 13,856 & 6,862 &
           1,428 & 41 & 
           1,576 \\
\quad \# of blocking bugs & 29,021 & 11,415 
           & 2,236 & 97 & 
           23,734\\
\midrule
Priority info &&&&& \\
\quad P1 & 6,737 & 1,165 & 47 & 0 & 517 \\
\quad P2 & 2,720 & 815 & 132 & 4 & 2,150\\
\quad P3 & 6,880 & 1,485 & 15,811 & 98 & 62,590\\
\quad P4 & 693 & 211 & 76 & 1 & 3,792\\
\quad P5 & 4,449 & 529 & 162 & 11 & 1,119\\
\quad Missing & 78,996 & 8,739 & 0 & 0 & 0\\
\midrule
Severity info &&&&& \\
\quad blocker & 204 & 64 & 169 & 1 & 494 \\
\quad critical & 3,782 & 360 & 308 & 1 & 2,919\\
\quad major & 4,556 & 325 & 1,104 & 9 & 5,885\\
\quad normal & 88,443 & 11,976 & 11,384 & 38 & 46,147\\
\quad minor & 2,426 & 167 & 753 & 3 & 4,763\\
\quad trivial & 1,019 & 52 & 214 & 1 & 1,366\\
\quad enhancement & 45 & 0 & 2,296 & 61 & 8,594\\
\midrule
Number of comments &&&&& \\
\quad mean & 8.1 & NA & 7.89 & NA & 8.5\\
\quad median & 4.0 & NA & 5.0 & NA & 6.0 \\
\quad standard deviation & 16.69 & NA & 9.6 & NA & 8.7\\
\bottomrule
\end{tabular}}
\end{table}

Priority comes from either the bug's assignee or the project lead. Generally, the bugs are triaged based on their priority, where P1 refers to the most significant bugs, whereas P5 corresponds to the least important bugs. The priority of bugs may change during the bug resolution process. For instance, when a developer observes that a bug takes excessive time to be solved, they assign a lower priority and start working on another one. We note that in Mozilla, 78.6\% of the bugs are not assigned a priority level; on the other hand, in Eclipse and LibreOffice, most of the bugs are assigned the medium level of P3, and the variation in priority is negligible. These observations are consistent with previous studies claiming that both ``priority'' and ``severity'' are unreliable factors~\citep{Shirin2020}.

The person who reports a bug (i.e., reporter) usually sets the severity to reflect how much it affects the user. To some extent, the reporter could overestimate this severity, and thus, it might need a revision from a developer. If users continually report bugs while assigning incorrect severity, they will damage their reputation and, in the long run, get less attention. Therefore, it is likely that a new user may tend to set the highest possible severity and make the severity level subjective. Bugzilla has a limit of ``Normal'' severity level for regular users, and the higher severity can be assigned only by contributors, developers, leaders, or admins.

The severity information is also used to differentiate between a bug and an enhancement report. Not all severity levels are accessible to regular users. Table~\ref{tab:bug_info} indicates that most of the bugs receive the ``Normal'' severity, the highest accessible level for ordinary users. Lastly, the number of comments below a bug report is an indicator of the engagement of users or developers in the bug solving process. The bug triage relies upon the bug comments; however, some noisy comments may affect this value~\citep{Xuan2012}. \textcolor{black}{Therefore, we do not use the number of comments in our prioritization or triage tasks.}



\section{Wayback Machine mechanism}\label{sec:Wayback}
Using the ITS information, we created a \textcolor{black}{past event regenerator that requires an} evolutionary database in which all bugs are sorted by their events' timestamp. The events include ``introduced'', ``resolved'', ``closed'', ``blocks'', ``depends on'', and ``reopened''. We ignore other events such as ``new'', ``verified'', or unimportant updates. Afterward, our event-based Wayback Machine gets updated whenever we have a new event in the system. If a user reports a new bug, it is added to the BDG with its full information retrieved from the Bugzilla ITS. If a bug blocks or depends on a new bug, we update the BDG by adding a new arc from the blocking bug to the blocked one. If a bug is resolved, we remove it from the BDG; however, we keep track of its information in a separate dataset, called the ``resolved dataset''. Using that, we can add back the bug to the BDG with its dependency information in the case of reopening.

As recalculating BDG information per event has a high complexity, we only update the information of the affected bugs. For instance, if a bug is linked to other bugs and is resolved in this timestamp, we update the depth and degree information of those bugs in the same subgraph. Using our Wayback Machine, we may retrieve the BDG information at any given time. Algorithm~\ref{alg:Wayback_Machine} shows how the ITS Wayback Machine works.

\begin{algorithm}[!ht]
	\SetKwData{Ev}{Evolutionary Database}\SetKwData{BDG}{BDG}\SetKwData{Solv}{Solved bugs tracker} \SetKwData{Resolved}{Resolved dataset} 
	\SetKwData{DB}{$\mathscr{DB}$}
	\KwData{\Ev with $K$ events, information of the bugs extracted from Bugzilla (\DB)}
	\KwResult{Daily monitoring of bug dependency graph evolution}
    initialization;\\
	\emph{\BDG = $\emptyset$}\\
	\emph{\Solv = $\emptyset$}\\
	\emph{\Resolved = $\emptyset$}\\
	Sort \Ev by the changes' timestamps
	\BlankLine
	\For{$i \in \{1,\hdots,K\}$}{
		\begin{algorithmic}
        	\IF{\Ev$[i][\text{`status'}] == \text{introduced}$}
        		\STATE Add bug info to \BDG using \DB
        		\STATE Start solving time of the bug
        	\ELSIF{\Ev$[i][\text{`status'}] \in \text{[blocks, depends on]}$}
        		\STATE Add a directed arc from blocking to blocked bug in \BDG
        	\ELSIF{\Ev$[i][\text{`status'}] == \text{resolved}$}
        		\STATE Remove the bug from \BDG and add it to \Resolved
        		\STATE Update solving time of the bug
        	\ELSIF{\Ev$[i][\text{`status'}] == \text{reopened}$}
        		\STATE Remove the bug from \Resolved and add it back to \BDG
        		\STATE Update solving time of the bug
        	\ENDIF
        	
		Update \Solv in case we have a reopened, resolved, or introduced bug. \\
		Update the graph information of the bugs that are affected by event $i$.
		\end{algorithmic}

	}
	\caption{Wayback Machine}
	\label{alg:Wayback_Machine}
\end{algorithm}

\begin{figure*}[!ht]
\centering
\centerline{\includegraphics[width=\linewidth]{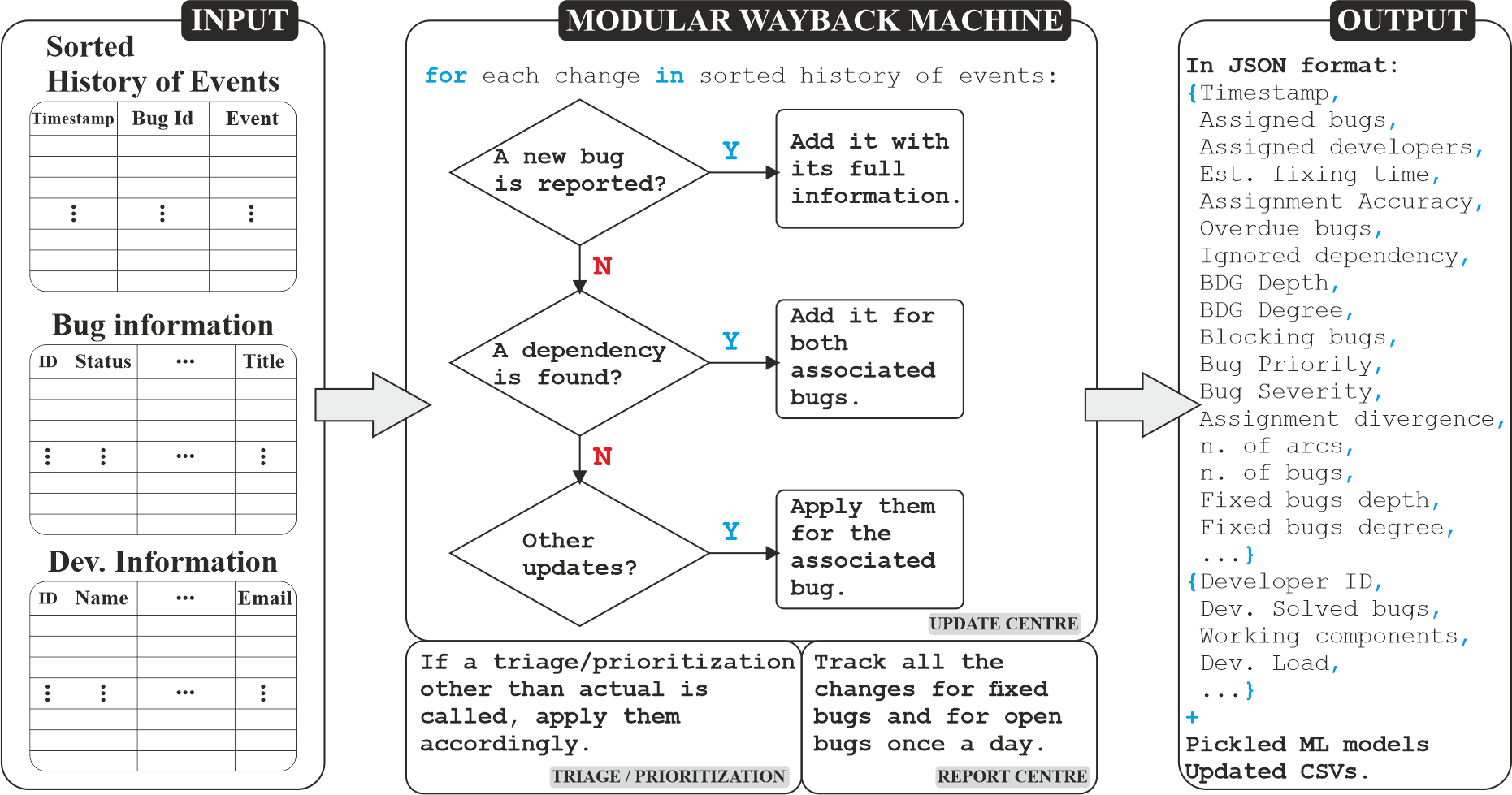}}
\caption{\textcolor{black}{The modular Wayback Machine with comprehensive reports as a way to evaluate different prioritization and triage algorithms.}}
\label{fig:simulation_scheme}
\end{figure*}

We model the actual bug tracking system via a discrete-event system simulation and explore the triage and prioritization decisions in the same environment. The timestamps of the bug reports and their dependency information are directly adopted from the ITS. \textcolor{black}{Therefore, the mechanism is more of a past-event regenerator than a simulator. 
\textcolor{black}{That is, this tool does not rely on a stochastic process or a fitted statistical distribution to an extracted dataset that would lead to randomly generated events.} 
The event regenerator, which we call Wayback Machine,} is run for all the reports between 2010 and 2020. \textcolor{black}{Figure~\ref{fig:simulation_scheme} illustrates a simplified version of the Wayback Machine together with its inputs and outputs. We sort the events by their chronological order. The events include new bug reports, blocking information of the bugs, assigning information, bug reopenings, bug resolution or closing time, and new comments to the system. The event list can be further expanded to include cc'ed people and changes in severity or priority of the bugs. Moreover, we separately utilize bugs' and developers' tabular information as the other two model inputs. The modular Wayback Machine comprises three segments, namely, the update centre, optional customized triage or prioritization module, and report centre.}

\textcolor{black}{In the update center, the actual historical events are run one at a time according to their timestamp.} At each timestamp $t$, we check if bug $b_i^d$ assigned to developer $d$ should be solved. If there exists any bug to be solved and no other bug blocks it, we fix and remove it from the BDG. 
If we do not have any bug to solve at this timestamp, we may update the system with a new blocking, reopening, closing, assigning, or fixing information. We also continue adding new bugs based on their actual report time which expands the BDG. 

\textcolor{black}{In the last module, we track all the changes that the whole ecosystem undergoes. By default, the report centre records the changes on a daily basis; however, the granularity of the record times can be manually changed by the user. The reports have three main parts: the major changes to the BDG, the detailed updates of the fixed and postponed bugs, and the schedule and the list of assigned bugs to the developers. These comprehensive metrics are recorded and presented as the output at the end of the testing phase.}

\section{Bug prioritization and triage problems}
\label{sec:BugPrioritizationTriage}
\textcolor{black}{Bug prioritization determines the priority of a bug according to its severity and urgency. On the other hand, although bug triage is relevant to bug prioritization, it also inspects a bug, understands its content, prioritizes, and finally assigns it to a proper developer using a variety of bug features~\cite{alazzam2020automatic, Hooimeijer2007}. In the Wayback Machine, we may use the optional triage or prioritization module and implement a new related algorithm; accordingly, the actual triage/assignment decisions will be substituted by the ones proposed by this module (see Figure~\ref{fig:simulation_scheme}). Hence, we may observe how the BDG evolves if we replace the actual assignment decisions with the proposed algorithm. As such, the Wayback Machine provides a practical perspective towards the performance of a suggested prioritization/triage model.}

In the triage process, we assume that developers cannot work on more than a bug at the same time. \textcolor{black}{Although this is a strong assumption, it is compatible with previous studies~\cite{Kashiwa2020, jahanshahi2021dabt} and is based on the fact that the exact schedules of the developers are not known with certainty. We also presume when we prioritize a bug over others based on an algorithm, we assign it to the most appropriate developer. Therefore, as we only investigate the prioritization accuracy and not the assignment accuracy in the prioritization task, we assume all bug assignments are done to the right developer. Nevertheless, in bug triage, the model decides on the assigned developer, and the above assumption only holds for the prioritization task.}

\subsection{Data preprocessing}
After collecting data and building the database, we implement the below steps to prepare the data for the Wayback Machine.
\begin{itemize}
    \item We remove duplicate bugs, and whenever a duplicate bug has more information than the original one (e.g., dependency information or general information), we merge its information with the original bug's information. \textcolor{black}{This is similar to developers' practice in the ITS and in line with the previous studies~\citep{Shirin2020}.} 
    \item Dependency information of older bugs is kept if and only if it \textcolor{black}{affects} the targeted bugs. 
    \item ``Enhancement'' reports are eliminated from the database as they do not represent a real defect in the system \textcolor{black}{(see~\citep{Shirin2020,Kashiwa2020}). We consider all the enhancements according to the bugs' last status.} 
    \item Some of the bugs were not accessible through REST API as a basic user. Hence, their information is not included.
    \item As there are many lingering bugs in the system that remain unresolved, we decided to disregard these cases since the bugs with an extraordinary fixing time are considered to be outliers in the system~\citep{jahanshahi2021dabt}.
\end{itemize}

\textcolor{black}{\paragraph{Feasible bug prioritization/triage cases} Not all the bugs are feasible to be assigned/prioritized. We clean the data step by step and report the result only for the feasible bugs. Feasible bugs should
\begin{itemize}
    \item have the resolved status by the end of 2019;
    \item be solved by active developers \textendash i.e., developers whose bug fix number is higher than the interquartile range (IQR) of bug fix numbers of all developers;
    \item have the exact assignment date (in some cases, the assignment date is not recorded in the history of the bugs, and we exclude those bugs);
    \item have acceptable fixing time \textendash i.e., their fixing time should be smaller than $\text{Q3} + (1.5 \times \text{IQR}) $, where $\text{Q3}$ and $\text{IQR}$ are the third quartile and interquartile range of the bug fixing time, respectively. 
\end{itemize}
We take the number of active developers as 28, 86, and 124 for EclipseJDT, LibreOffice, and Mozilla, respectively. The cleaning process is similar to that of~\citet{Kashiwa2020} and \citet{jahanshahi2021dabt}.}

\subsection{Performance metrics} \label{sec:perf_metrics}
\textcolor{black}{We define various metrics to compare different prioritization and triage strategies. These metrics include static metrics (e.g., assignment accuracy) and evolutionary metrics (e.g., percentage of overdue bugs). Note that evolutionary metrics cannot be easily reported unless a Wayback Machine is used. That is, these are related to the time when a bug is assigned or prioritized while considering either the developers' workload at the assignment time or the status of other open bugs in the system. To incorporate the impact of a triage or prioritization decision, we require considering time-related measures as well. The complete set of metrics that we used in our experiments with various bug prioritization and triage strategies are as follows.}

\begin{itemize}
    \item \textcolor{black}{\textbf{The Number of Assigned Bugs} represents the total number of assigned bugs during the testing phase. In practice, developers attempt to keep the number of open bugs in the system as low as possible. Therefore, they assign higher priority to the bugs that are more critical and/or easier and/or faster to be solved. The number of assigned bugs consists of the feasible bugs assigned by a specific method during the testing period~\citep{Shirin2020, Kashiwa2020}.}
    
    \item \textcolor{black}{\textbf{(Early, On-time, Late) Prioritization} indicates how many of the prioritized bugs are early, on-time, or late compared to actual assignments. It shows whether a prioritization strategy follows a similar pattern as the actual case.}
    
    \item \textcolor{black}{\textbf{Assigning Time Divergence}, similar to previous metrics, shows the standard deviation of the prioritization times compared to the actual case. The smaller value for the metric is desirable.}
    
    \item \textcolor{black}{\textbf{Mean Fixing Time} illustrates the average fixing time of a bug. As the fixing time of a bug is defined based on the developer to whom it is assigned, this factor shows how a triage algorithm considers fixing time~\citep{Kashiwa2020}.}
    
    \item \textcolor{black}{\textbf{The Number of Assigned Developers} is of importance as it can be useful to see how many developers are selected by a triage algorithm during the testing phase~\citep{Kashiwa2020}.}
    
    \item \textcolor{black}{\textbf{Task Concentration} among developers shows how fair is the assignment distribution among them. Previous studies~\cite{park2011costriage, Kashiwa2020} indicate that some algorithms overspecialize, i.e., they assign all the bugs to few expert developers. Therefore, smaller task concentration shows a better distribution among developers.}
    
    \item \textcolor{black}{\textbf{Assignment Accuracy} is an important measure as it helps understanding how a triage algorithm mimics the actual case. An accurate assignment is defined as assigning bug $b$ from component $c$ to developer $d$, who has previous experience in addressing bugs of type $c$~\citep{park2011costriage, mani2019deeptriage}.}
    
    \item \textcolor{black}{\textbf{Percentage of Overdue Bugs} determines how many bugs cannot be fixed before the next release. This metric can be computed only if we regenerate past events. If we assign more bugs to a developer than they can handle, those bugs will more likely get overdue. Therefore, using proper timing in the assignment is necessary~\citep{Kashiwa2020, jahanshahi2021dabt}.}
    
    \item \textcolor{black}{\textbf{Infeasible Assignment with respect to the BDG} shows the percentage of the assigned bugs that had a blocking bug. These are infeasible assignments and need to be postponed until the parent bug is resolved. This evolutionary metric also requires the Wayback Machine as it relies on the information related to the time when a dependency is found and the fixing time of the blocking bug~\citep{jahanshahi2021dabt}.}
\end{itemize}

\subsection{Bug prioritization strategies} \label{sec:priori_strategies}
In practice, triagers may use a combination of factors, such as validity, reproducibility, severity, priority, and even customer pressure to choose an appropriate bug to fix. In some cases, they may also decide based on the blocking effect of a bug. \textcolor{black}{Thus, we define a list of prioritization strategies, including the graph-related (i.e., using features coming from the BDG), severity- and priority-based, as well as machine learning-based approaches. 
Any other prioritization strategy can be added to the modular Wayback Machine in a similar manner. 
\textcolor{black}{On the other hand, we note that bug prioritization and triage is a multifaceted problem with many human factors at play. Triaging and prioritization policies may exist at any organization, but developers may deviate from those to some extent, often for good reasons. Hence, we acknowledge that a naive rule-based approach or even a machine learning algorithm may not be able to fully cover different intricate aspects of the bug prioritization task in practice.}}
\begin{enumerate}
    \item \textcolor{black}{ \textbf{Maximum sum of degree and depth}: This strategy selects the bug with the highest sum of its degree and depth. We take ``degree'' as the out-degree of a bug. Also, the depth of a bug in a directed graph is the maximum shortest path from the bug to any other bugs in the graph. \citet{Shirin2020} take this as a potential, unbiased factor in bug prioritization.}
    
    \item \textcolor{black}{\textbf{Maximum priority}: This rule-based strategy chooses the bug that has the highest priority among other open bugs. In case of ties, it chooses one high-priority bug arbitrarily. Therefore, we repeat the experiment with this strategy and take the average performance. As we explored the importance of priority in RQ1b, we decide to keep it as an option to examine its similarity to the actual case.}
    
    \item \textbf{Maximum severity}: This strategy chooses bugs with the highest severity first. This approach might be controversial due to the lack of objective assessment of the severity scores; however, we keep this strategy as an alternative approach to the existing ones\textcolor{black}{, as discussed in RQ1b.}
    
    \item \textcolor{black}{\textbf{Cost-oriented strategy}: It computes the fixing time of a bug based on the Latent Dirichlet Allocation (LDA) similar to that of \citet{park2011costriage}. Specifically, we cluster bugs using the LDA algorithm and compute the average bug fixing times per topic/cluster. Accordingly, we prioritize the bugs that have the least estimated fixing time, i.e., cost.}
    
    \item \textcolor{black}{\textbf{Estimated Priority}: We predict the priority using support vector machine (SVM) after converting the textual information of the bug to numeric values using TF-IDF~\citep{kanwal2012bug}. We train our model on the TF-IDF output of bugs' titles and descriptions given their current priority levels. Accordingly, given a new bug report, the model can predict its priority level. The bugs with the highest estimated priority are selected at each timestamp.}
    
    \item \textcolor{black}{\textbf{Cost and Priority Consideration}: We combine cost- and priority-based strategies as follows. First, we normalize the estimated fixing time $c_i$ and estimated priority $p_i$ of bug $i$ to the range of 1 to 5. Then, we choose the bugs based on the below formula:}
    \begin{equation*}
        \big(\alpha \cdot \frac{p_i}{\max_i\{p_i\}}\big) + \big( (1-\alpha) \cdot \frac{\nicefrac{1}{c_i}}{\nicefrac{1}{\min_i\{c_i\}}} \big).
    \end{equation*}
    \textcolor{black}{We set the default value for the control parameter $\alpha$ to 0.5, which assigns identical importance to the priority and fixing cost. The bug with the highest aggregate value will be selected.}
    
    \item \textbf{Random}: This approach is considered as a naive baseline and corresponds to selecting the candidate bug randomly. We use this strategy to show how well other strategies perform compared to a random selection\textcolor{black}{, which is not recommended for bug prioritization tasks in practice.}
\end{enumerate}

\subsection{Bug triage strategies} \label{sec:triage_strategies}
\textcolor{black}{While prioritization techniques explore the order in which the bugs should be addressed, in the triage process we also consider the assignment of the bugs to proper developers in a timely manner. We evaluate different well-established bug triage algorithms, together with the actual case. However, as the Wayback Machine is a modular past-event regenerator, any other triage algorithm can be applied in the same context and be compared with these baselines. The source code and all datasets are available on GitHub\footnote{\url{https://github.com/HadiJahanshahi/WaybackMachine}}.}

\begin{enumerate}
    \item \textcolor{black}{\textbf{CBR}: Content-Based Recommendation (CBR) aims to assign a bug to the most appropriate developer through analyzing its content, i.e., its summary and description~\citep{anvik2006should}. This method converts bug titles and descriptions to numeric vectors and uses assigned developers as the labels. Previous studies show that SVM has the best performance for this classification task, and hence, we use the same approach in our analysis~\citep{anvik2006should, Lin2009}.}
    
    \item \textcolor{black}{\textbf{DeepTriage}: DeepTriage is developed based on the observation that bag-of-words of TF-IDF as a feature representation is unable to capture the semantic of the text and loses the order of the words~\citep{mani2019deeptriage}. Therefore, using a deep learning algorithm together with a word embedding, e.g., word2vec or paragraph vector can alleviate such issues. Accordingly, we re-implement the algorithm using Wayback Machine and report its performance through our novel evolutionary metrics.}
    
    \item \textcolor{black}{\textbf{CosTriage}: In the cost-aware recommendation system, not only the accuracy of the assignment but also its fixing cost is of importance~\citep{park2011costriage}. Accordingly, it combines CBR with a collaborative filtering recommender (CF) and builds developer profiles to estimate the approximate fixing time of each bug type. Bug types are determined by the LDA using summary and description. The trade-off between accuracy and fixing time can be formulated as
    \begin{equation*}
        \big(\alpha \frac{s_i^d}{\max_d\{s_i^d\}}\big) + \big( (1-\alpha)  \frac{\nicefrac{1}{c_i^d}}{\nicefrac{1}{\min_d\{c_i^d\}}} \big),
    \end{equation*}
    where $s_i^d$ is the suitability of bug $i$ when assigned to developer $d$, $c_i^d$ is the estimated fixing time coming from the CF for bug $i$ when assigned to developer $d$, and $\alpha$ is a control parameter~\citep{park2011costriage}. The suitability is estimated by the SVM similar to CBR. In this study, we set the value of 0.5 for $\alpha$; however, the Wayback Machine can dynamically change it.}
    
    \item \textcolor{black}{\textbf{Random}: This naive strategy randomly assigns a candidate bug to a developer. While using this strategy, we repeat the experiment 5 times and report the average performance. We acknowledge that a naive rule-based approach cannot address the bug triage task. We utilize it only as a baseline, and it is not a \textcolor{black}{recommended way to address bug triaging}.}
\end{enumerate}

\section{\textcolor{black}{Results}}\label{sec:findings}
\textcolor{black}{We evaluate the proposed Wayback Machine in two ways. First, we investigate the ability of the past-event regenerator to provide practical information related to past prioritization and triage decisions. It includes exploring the number of bugs and dependencies, together with the depth, degree, severity, and priority of the open bugs \textcolor{black}{compared} to the fixed bugs over time. Second, we assess the ability of the Wayback Machine in incorporating prioritization and triage algorithms. We report the performance of these algorithms considering the evolutionary nature of the ITS. \textcolor{black}{We also contacted active developers from the three projects under study to validate our findings. The study conclusions have been tailored according to their feedback.} }

\subsection{Evaluating the history of the ITS}\label{sec:RQ1}
\textcolor{black}{In this section, we present the results of our empirical study that answer two main research questions. 
For this purpose, we analyze the evolution of the bugs in the ITS and explore the effect of different bug prioritization and triage strategies. 
We characterize the bug dependency and its impact on lingering bugs during the evolution of three open-source software systems. We further investigate the actual evolutionary performance of well-established bug prioritization and triage strategies using the Wayback Machine.}

\begin{RQquestion}
    \textbf{RQ1a: How do open-source software systems evolve in terms of the number of bug reports, bug dependencies, and lingering bugs?}
\end{RQquestion} 

\begin{figure*}[!ht]
  \centering
  \medskip
  \begin{subfigure}[t]{.32\textwidth}
    \centering\includegraphics[width=\textwidth]{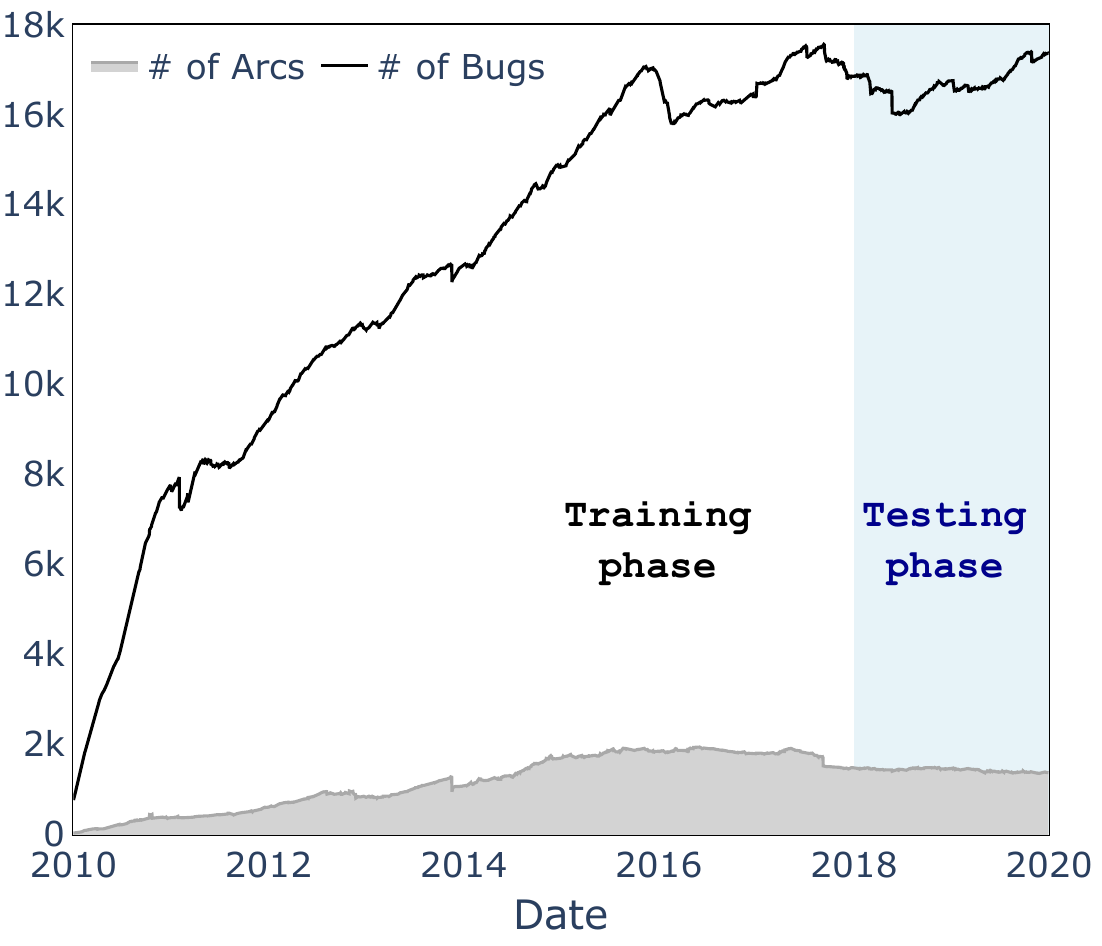}
    \caption{Mozilla} \label{fig:Mozilla_num}
  \end{subfigure}
  \begin{subfigure}[t]{.32\textwidth}
    \centering\includegraphics[width=\textwidth]{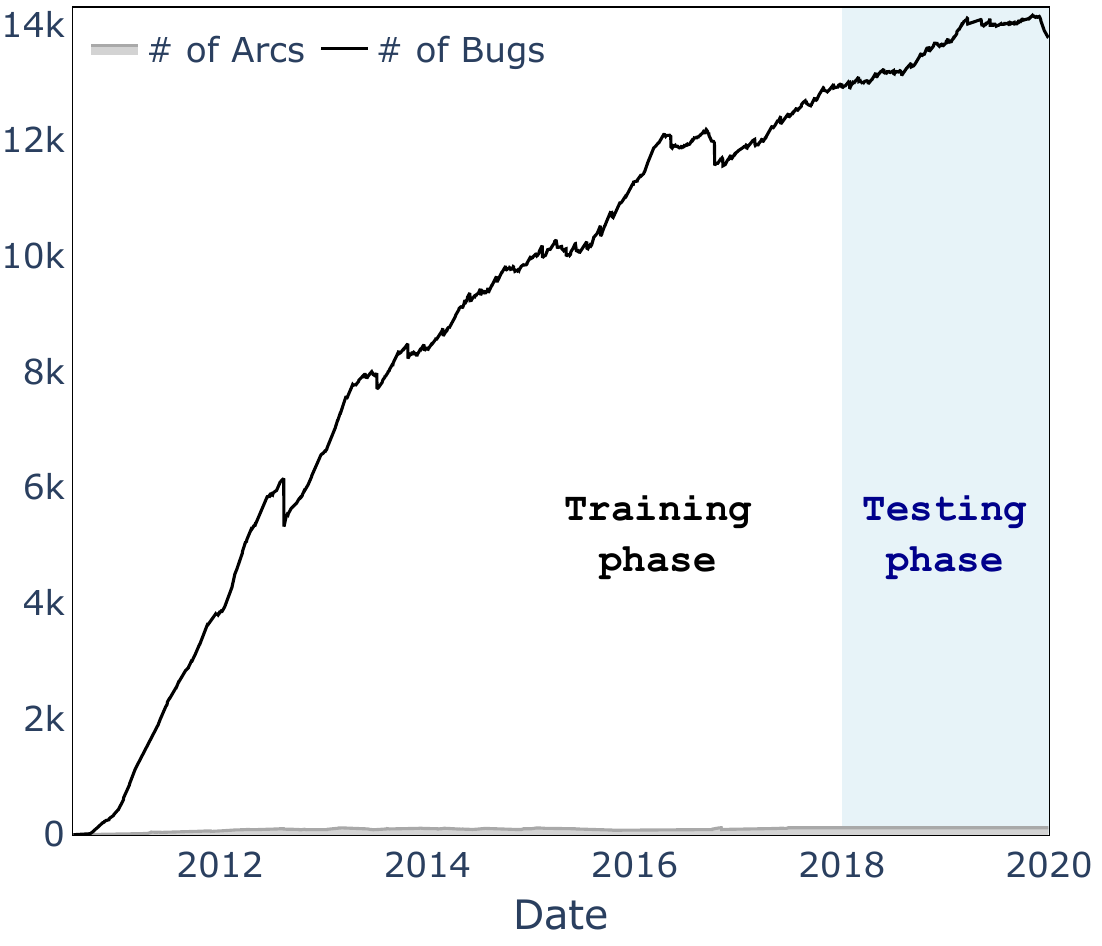}
    \caption{LibreOffice} \label{fig:LibOffice_num}
  \end{subfigure} 
  \begin{subfigure}[t]{.32\textwidth}
    \centering\includegraphics[width=\textwidth]{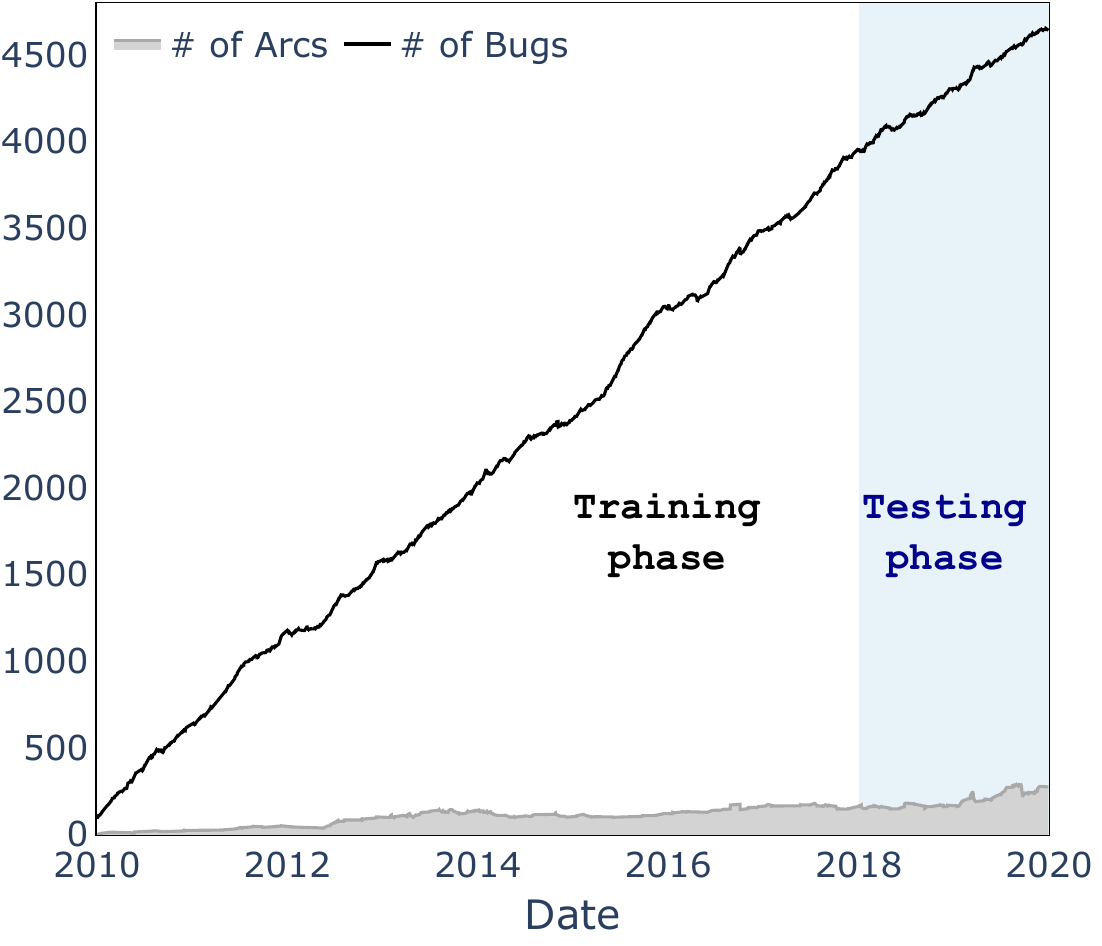}
    \caption{Eclipse} \label{fig:Eclipse_num}
  \end{subfigure}
    ~\caption{The number of nodes and arcs in bug dependency graph for Mozilla, LibreOffice, and Eclipse projects \textit{($x$-axis corresponds to the year and $y$-axis corresponds to the monthly bug and dependency counts; $y$-axis range differs for each project.)}.}
    \label{fig:number_of_bugs_and_arcs}
\end{figure*}

\textcolor{black}{The line plot in Figure~\ref{fig:number_of_bugs_and_arcs} shows the actual number of bugs, and the area plot shows the number of arcs (i.e., bug dependency) in each project during the last decade}. We extract dependencies from the bug's history and use the exact date when the dependency is determined. We observe significant differences between the projects. \textcolor{black}{The Eclipse JDT (Figure~\ref{fig:LibOffice_num}) has the lowest number of arcs among these projects. In this graph, we exclude meta bugs \textendash i.e., tracking bugs used to associate reports with useful data. We note that LibreOffice has very few reported dependencies. In fact, our interviews with LibreOffice developers confirmed this observation, where they noted that dependencies are not as frequently reported in LibreOffice as it is done in other projects. Therefore, bug dependency, in the case of LibreOffice, becomes a less important factor in triage and prioritization decisions. Developers in Mozilla (Figure~\ref{fig:Mozilla_num}) record the bug dependency during the project lifespan. Therefore, in the following research question, we investigate whether these dependencies influence the bug prioritization/triage process.}  

In the last period, the ratios of open bugs to the number of bug reports are $15\%$, $20\%$, and $28\%$ for Mozilla, LibreOffice, and Eclipse, respectively, which suggests a significantly higher rate of lingering bugs in the Eclipse project. Although Eclipse has only 16,342 bug reports, it contains 4,643 unresolved reports at the end of the period. This observation indicates that the number of arcs is not the only factor in lingering bugs. That is, there might be a shortage of developers, or the bugs in the Eclipse project might require more time to be resolved, or there might be a higher number of fastidious contributors reporting bugs that are less important and can be postponed.

\begin{figure*}[!ht]
  \centering
  \medskip
  \begin{subfigure}[t]{.32\textwidth}
    \centering\includegraphics[width=\textwidth]{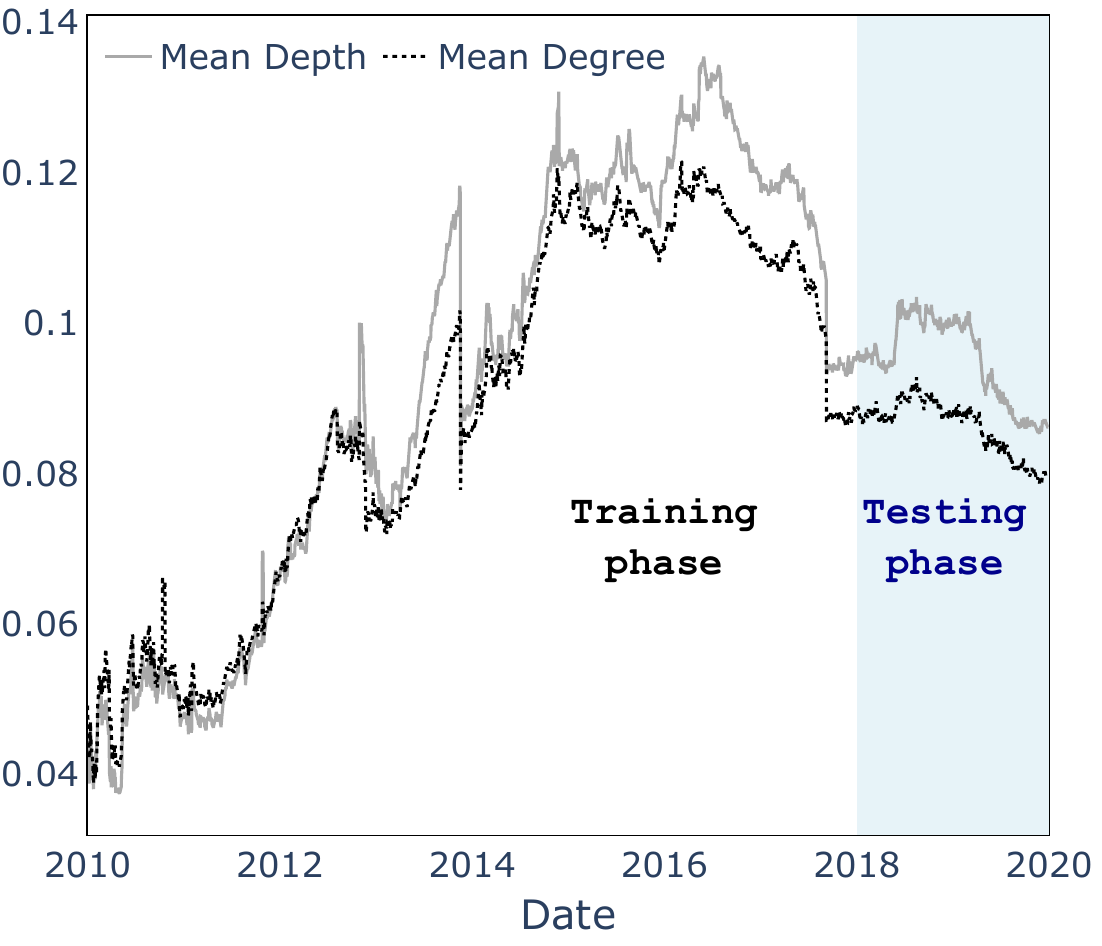}
    \caption{Mozilla} \label{fig:Mozilla_depth_degree}
  \end{subfigure}
  \begin{subfigure}[t]{.32\textwidth}
    \centering\includegraphics[width=\textwidth]{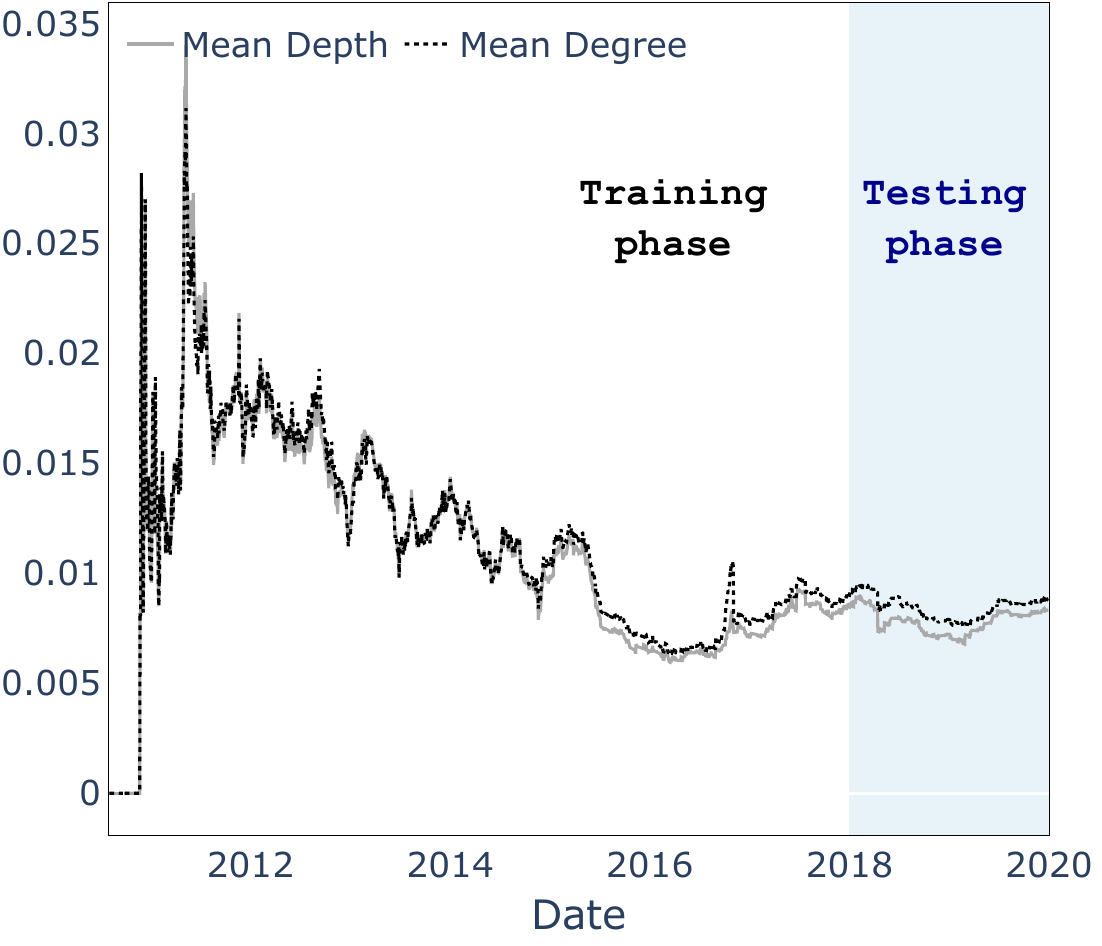}
    \caption{LibreOffice} \label{fig:LibOffice_depth_degree}
  \end{subfigure} 
  \begin{subfigure}[t]{.32\textwidth}
    \centering\includegraphics[width=\textwidth]{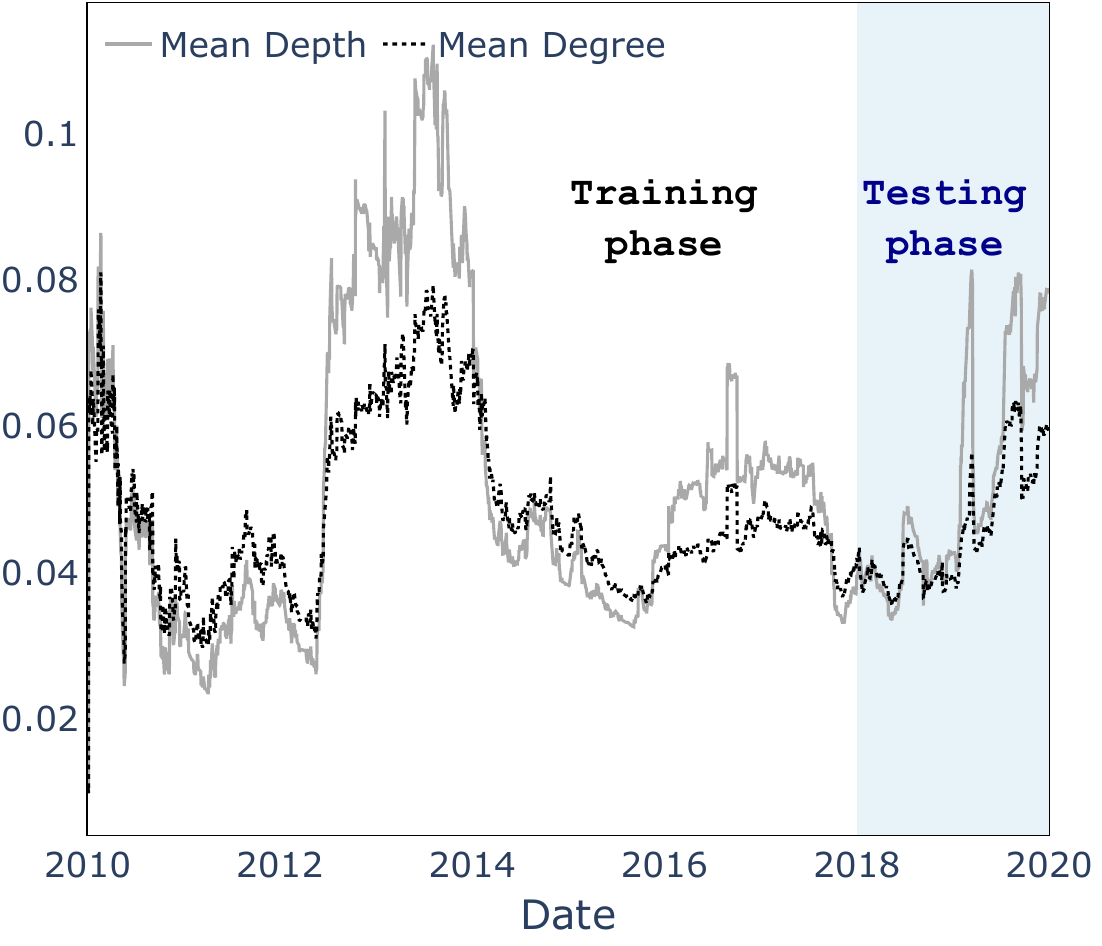}
    \caption{Eclipse} \label{fig:Eclipse_depth_degree}
  \end{subfigure}
    ~\caption{The monthly evolution of mean depth and degree of BDG for Mozilla, LibreOffice, and Eclipse projects \textit{($x$-axis corresponds to the year and $y$-axis corresponds to the mean depth and degree; $y$-axis range differs for each project.)}.}
    \label{fig:depth_degree}
\end{figure*}

Figure~\ref{fig:depth_degree} shows the degree and depth evolution of all three projects. \textcolor{black}{In the atypical case of LibreOffice, we observe that after the initial spike in the depth and degree of the bugs, they become stable and approach the value of 0.01 after 2015. Also, the average depth and degree are much smaller in LibreOffice, as shown by Figure~\ref{fig:LibOffice_num}. After 2017, developers in LibreOffice introduced a large number of meta bugs; however, we ignored these bugs as they are not real blocking bugs and rather act as a clustering approach to group similar bugs. On the other hand, the general trend of the degree and depth of the bugs in the Mozilla project is ascending until 2016 and then descending afterward, whereas those for the Eclipse project remain almost at the same level with some seasonal fluctuation. Therefore, we conclude that in terms of graph complexity, each project has its own characteristics that cannot be generalized to other cases.}

\begin{RQquestion}
    \textbf{RQ1b: How do the characteristics of the resolved bugs change over time?}
\end{RQquestion}

To address this research question, we compare the characteristics of the resolved bugs and open bugs to infer the notion behind the actual bug prioritization process. We are mainly interested in graph-related indices (e.g., degree and depth of the bugs) \textcolor{black}{together with severity and priority}. While comparing the actual decisions over time, we explore whether bug triagers consider dependency information, priority, and severity in bug prioritization. \textcolor{black}{Our main focus is the testing phase \textendash from 2018 to 2020. We assume that a triaged/fixed bug has a higher priority over deferred/unresolved bugs.}

\begin{figure*}[!ht]
  \centering
  \medskip
  \begin{subfigure}[t]{.31\textwidth}
    \centering\includegraphics[width=\textwidth]{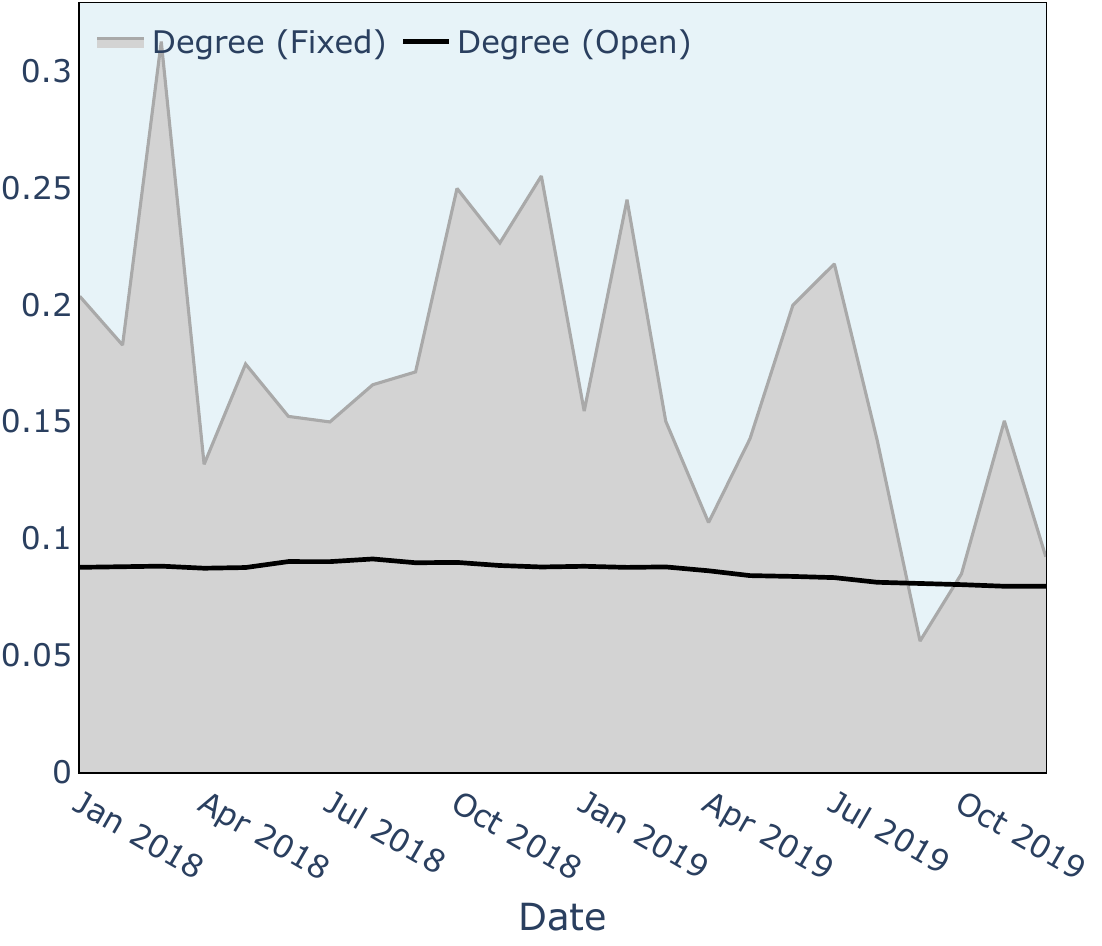}
    \caption{Mozilla (degree)} \label{fig:Mozilla_deg_solv}
  \end{subfigure}\quad
  \begin{subfigure}[t]{.31\textwidth}
    \centering\includegraphics[width=\textwidth]{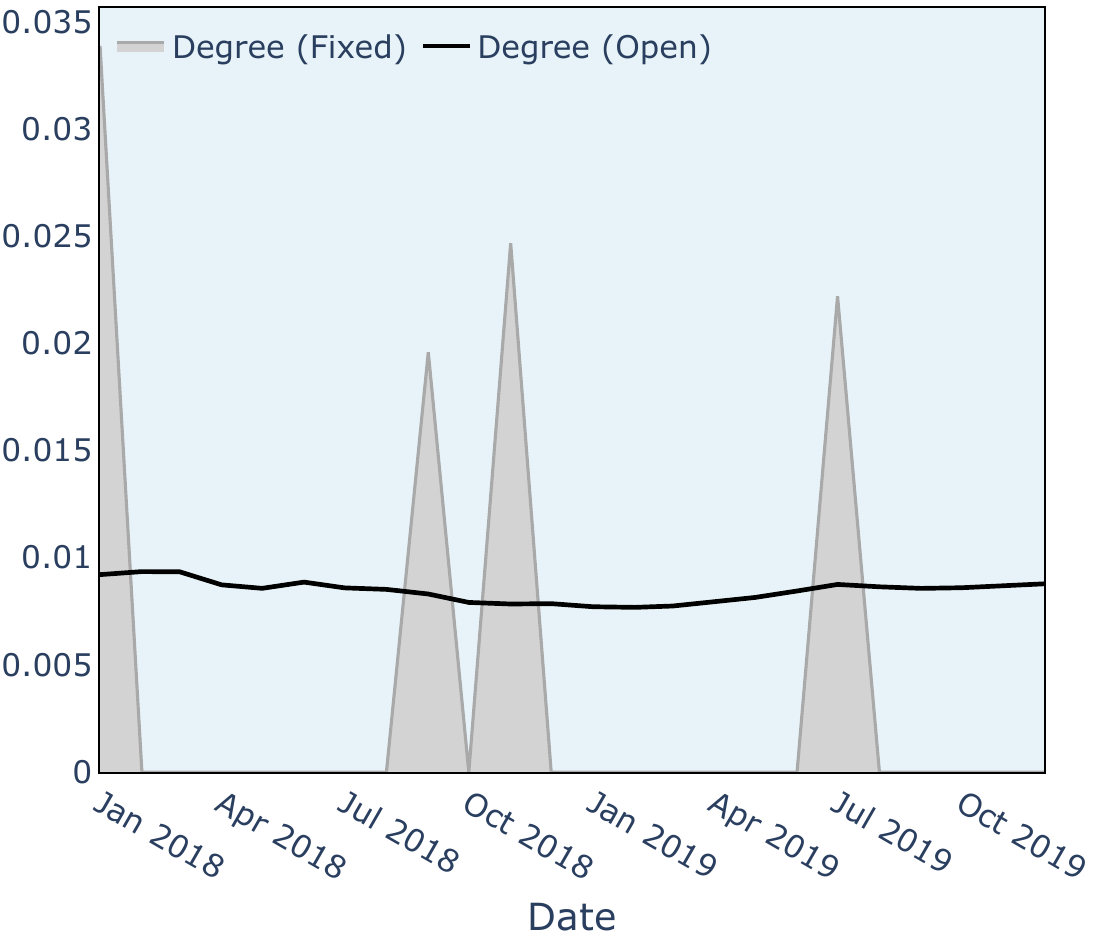}
    \caption{LibreOffice (degree)} \label{fig:LibOffice_deg_solv}
  \end{subfigure} \quad
  \begin{subfigure}[t]{.31\textwidth}
    \centering\includegraphics[width=\textwidth]{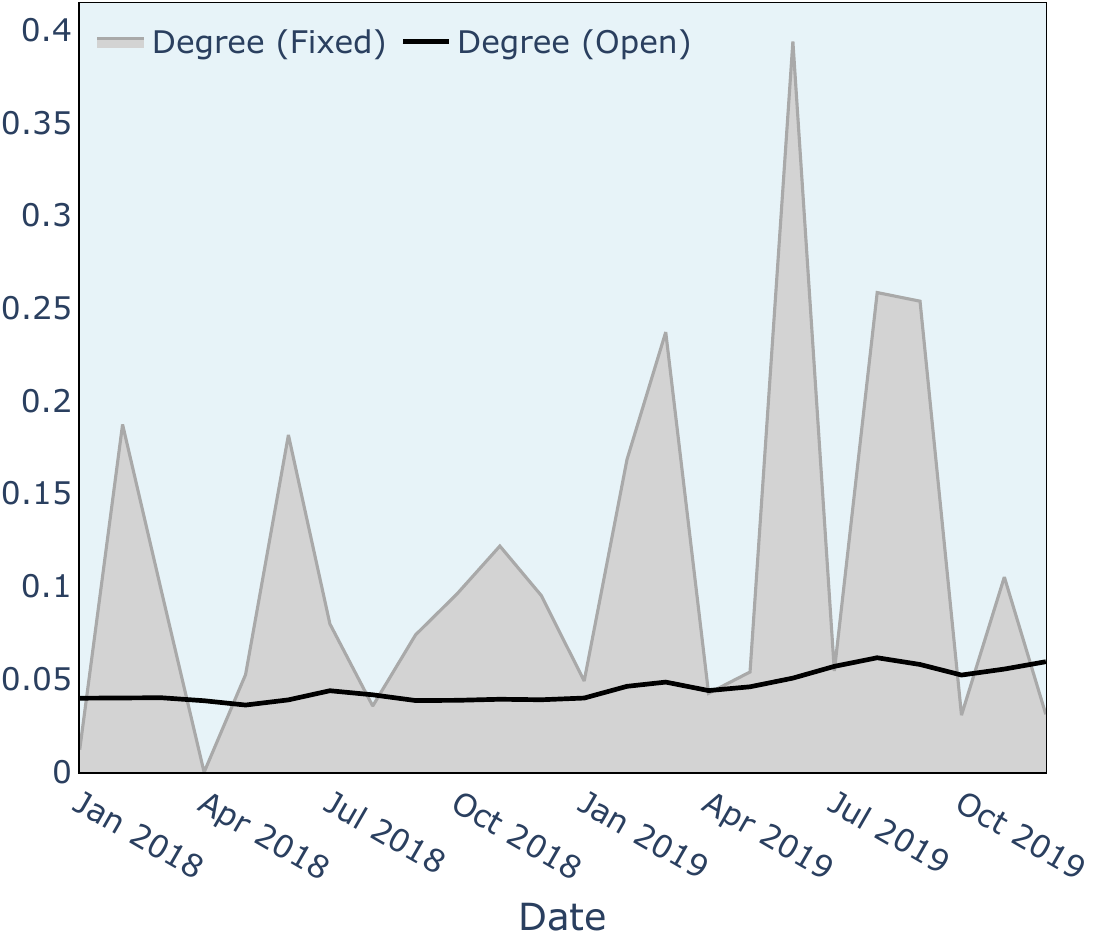}
    \caption{Eclipse (degree)} \label{fig:Eclipse_deg_solv}
  \end{subfigure} \\ 
  \medskip
  \begin{subfigure}[t]{.31\textwidth}
    \centering\includegraphics[width=\textwidth]{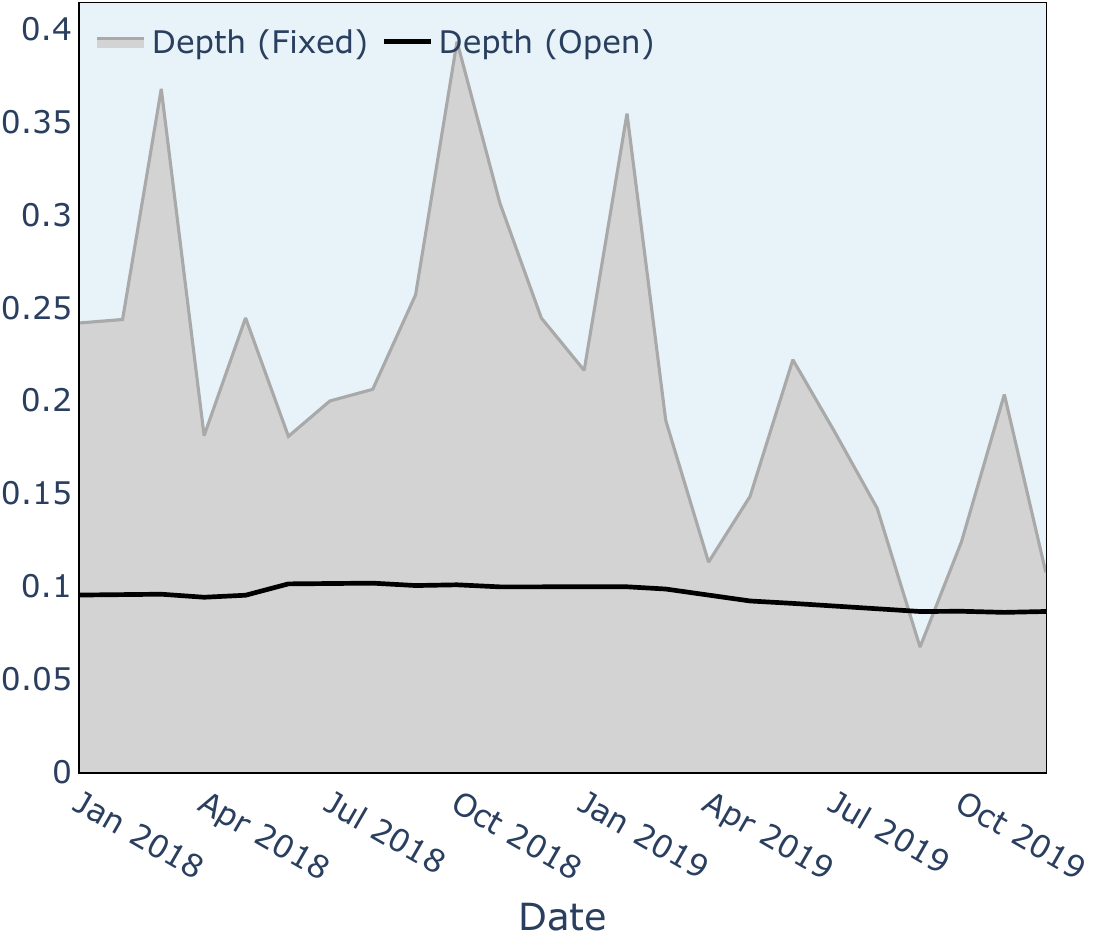}
    \caption{Mozilla (depth)} \label{fig:Mozilla_depth_solv}
  \end{subfigure}\quad
  \begin{subfigure}[t]{.31\textwidth}
    \centering\includegraphics[width=\textwidth]{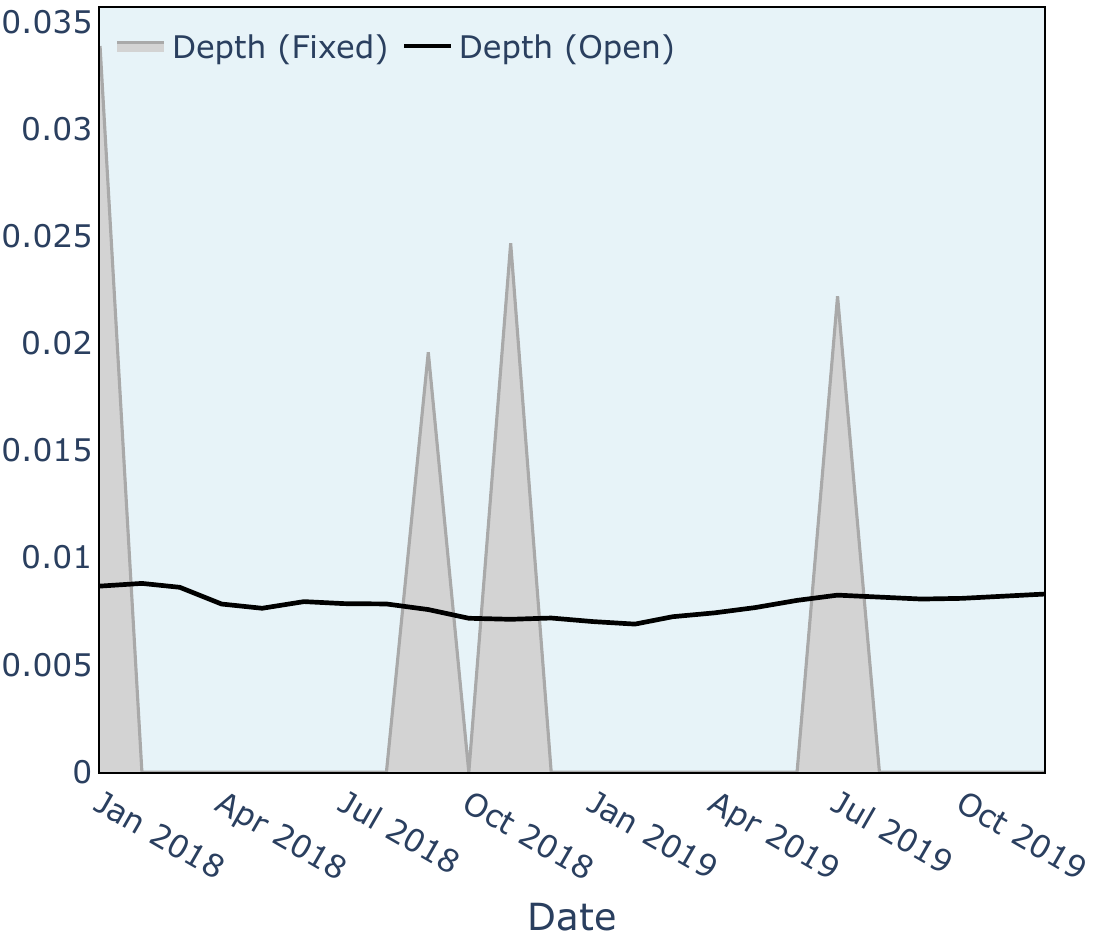}
    \caption{LibreOffice (depth)} \label{fig:LibOffice_depth_solv}
  \end{subfigure} \quad
  \begin{subfigure}[t]{.31\textwidth}
    \centering\includegraphics[width=\textwidth]{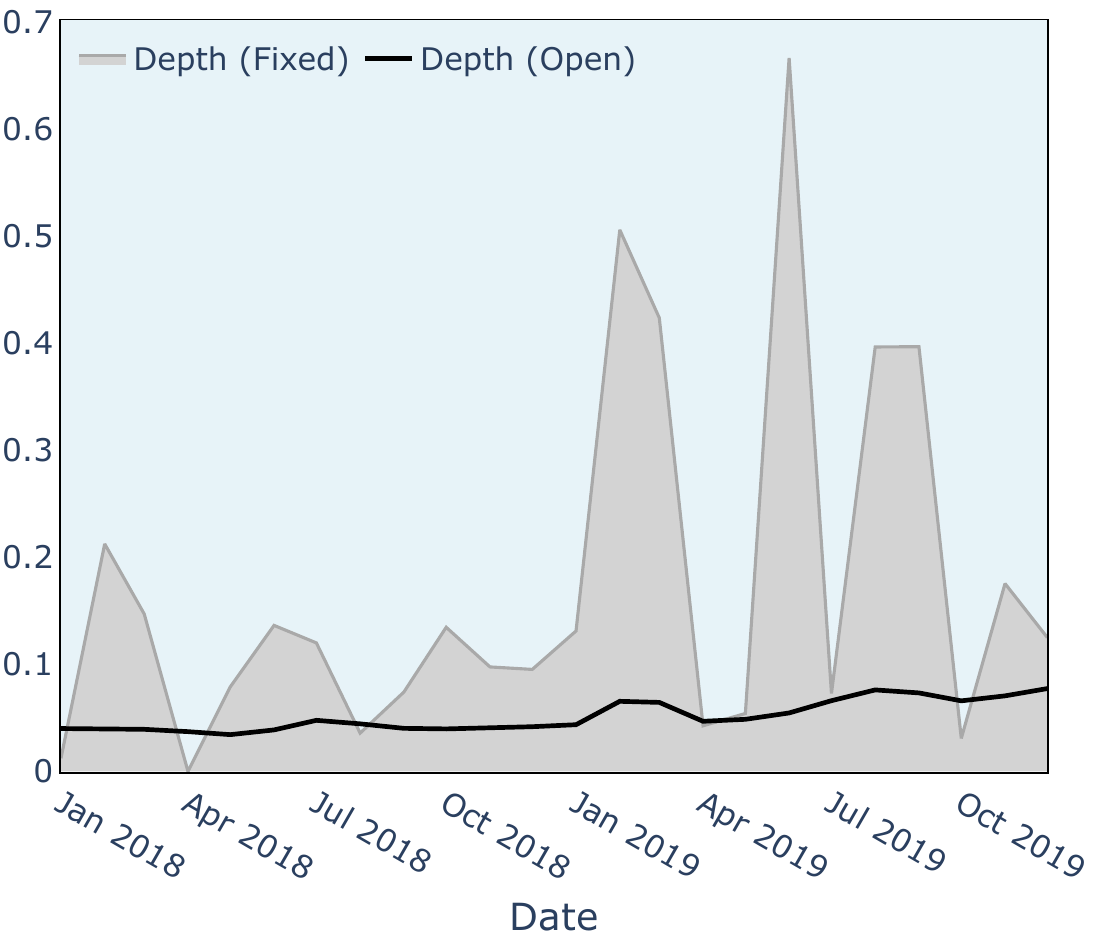}
    \caption{Eclipse (depth)} \label{fig:Eclipse_depth_solv}
  \end{subfigure}
    ~\caption{The comparison of the monthly depth and degree of the bugs in BDG and fixed bugs \textit{(the area plot shows the degree/depth of fixed bugs, whereas blue lines indicate the degree/depth of remaining bugs in the graph; $y$-axis range differs for each project.)}.}
    \label{fig:degree_of_solv}
\end{figure*}

Figure~\ref{fig:degree_of_solv} juxtaposes the degree and depth of the bugs that are solved with those of postponed bugs (i.e., remained open). Such a comparison provides a clear picture \textcolor{black}{of whether bug triagers prioritize a bug based on their dependency.} We show the average degree of the fixed bugs as an area plot and the average degree of the open bugs as a line graph. If we take the area plot as an upper bound of the line plot, we may conclude that, on average, the triagers prioritize the bugs with a higher degree. In Figures~\ref{fig:Mozilla_deg_solv} and \ref{fig:Eclipse_deg_solv}, \textcolor{black}{the grey region almost always contains the black line, meaning that, on average, the degree of solved bugs is greater than that of the postponed bugs. We use a one-tailed paired t-test with a significance level of 0.05 to check the validity of our observation. The null hypothesis is that the true degree/depth mean difference for fixed and unfixed bugs is equal to zero. For both projects, with a $p$-value of $4.3e-10$, we reject the null hypothesis.} Hence, triagers indirectly consider the dependency while addressing open bugs. In the special case of LibreOffice, where the BDG is very sparse (Figure~\ref{fig:LibOffice_deg_solv}), 
\textcolor{black}{we do not observe such behavior. The area plot is almost always zero, meaning that the blocking effect is not considered to be an important factor here.}

Regarding the average depth of fixed and open bugs, in Mozilla and Eclipse projects, the depth of the open bugs is mainly smaller than that of fixed bugs --i.e., the black line is within the area under the grey curve. \textcolor{black}{We also observe a similar behavior of the LibreOffice project as we explained for its degree. Our conclusion remains identical. The blocking bugs become important if and only if the blocking information is constantly recorded and the BDG is not sparse. We do not see any direct relationship with lingering bugs in this case. We find that in automating bug triage and bug prioritization process, researchers consider dependency together with other bug attributes. Prioritization based only on the bug dependency cannot be generalized~\cite{Shirin2020}}. 

\begin{figure*}[!ht]
  \centering
  \medskip
  \begin{subfigure}[t]{.31\textwidth}
    \centering\includegraphics[width=\textwidth]{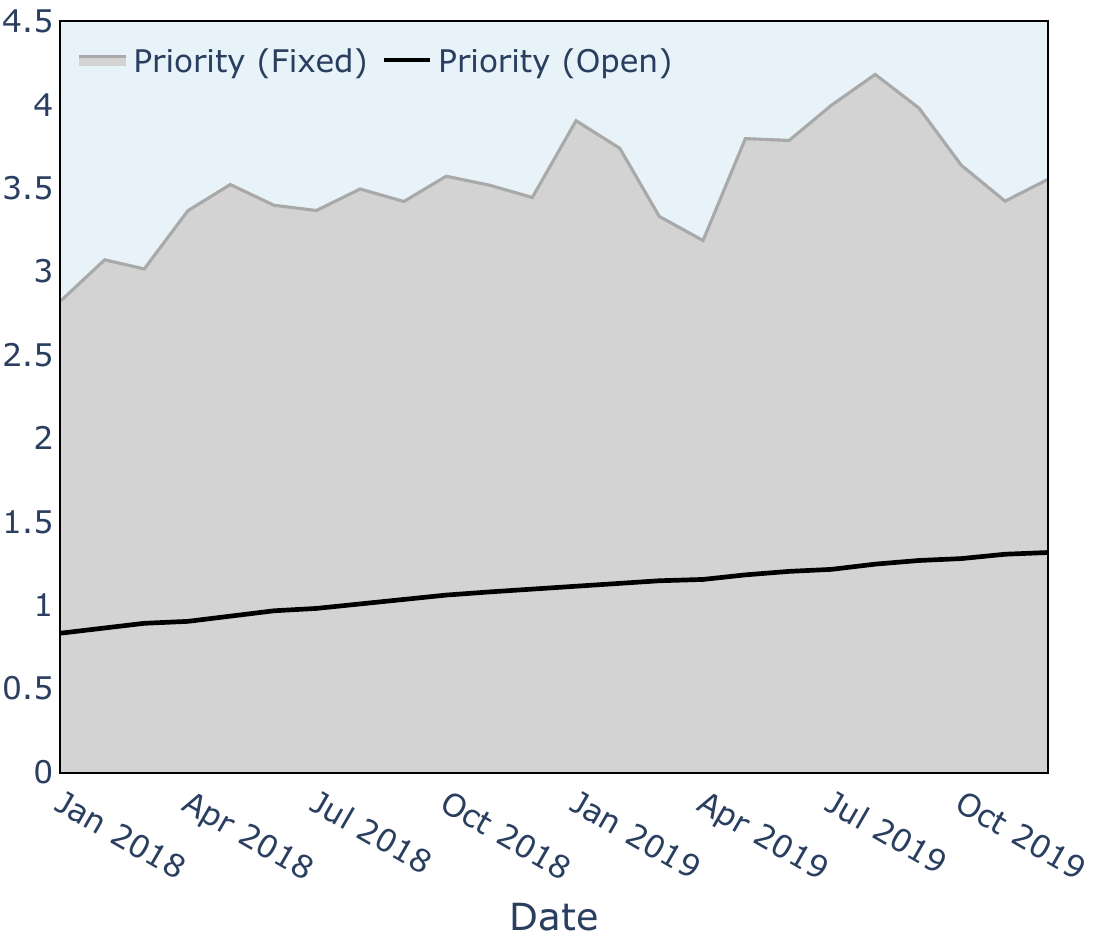}
    \caption{Mozilla (priority)} \label{fig:Mozilla_prio_solv}
  \end{subfigure}\quad
  \begin{subfigure}[t]{.31\textwidth}
    \centering\includegraphics[width=\textwidth]{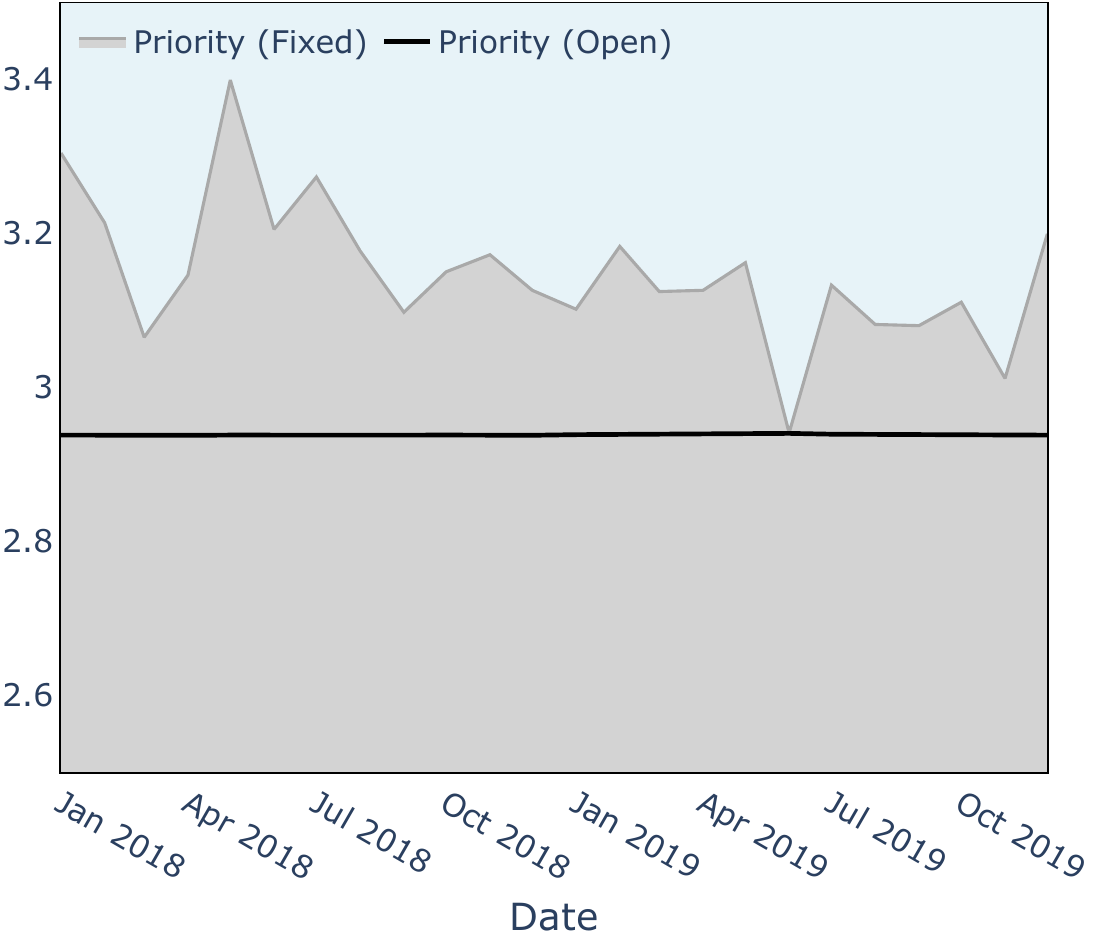}
    \caption{LibreOffice (priority)} \label{fig:LibOffice_prio_solv}
  \end{subfigure} \quad
  \begin{subfigure}[t]{.31\textwidth}
    \centering\includegraphics[width=\textwidth]{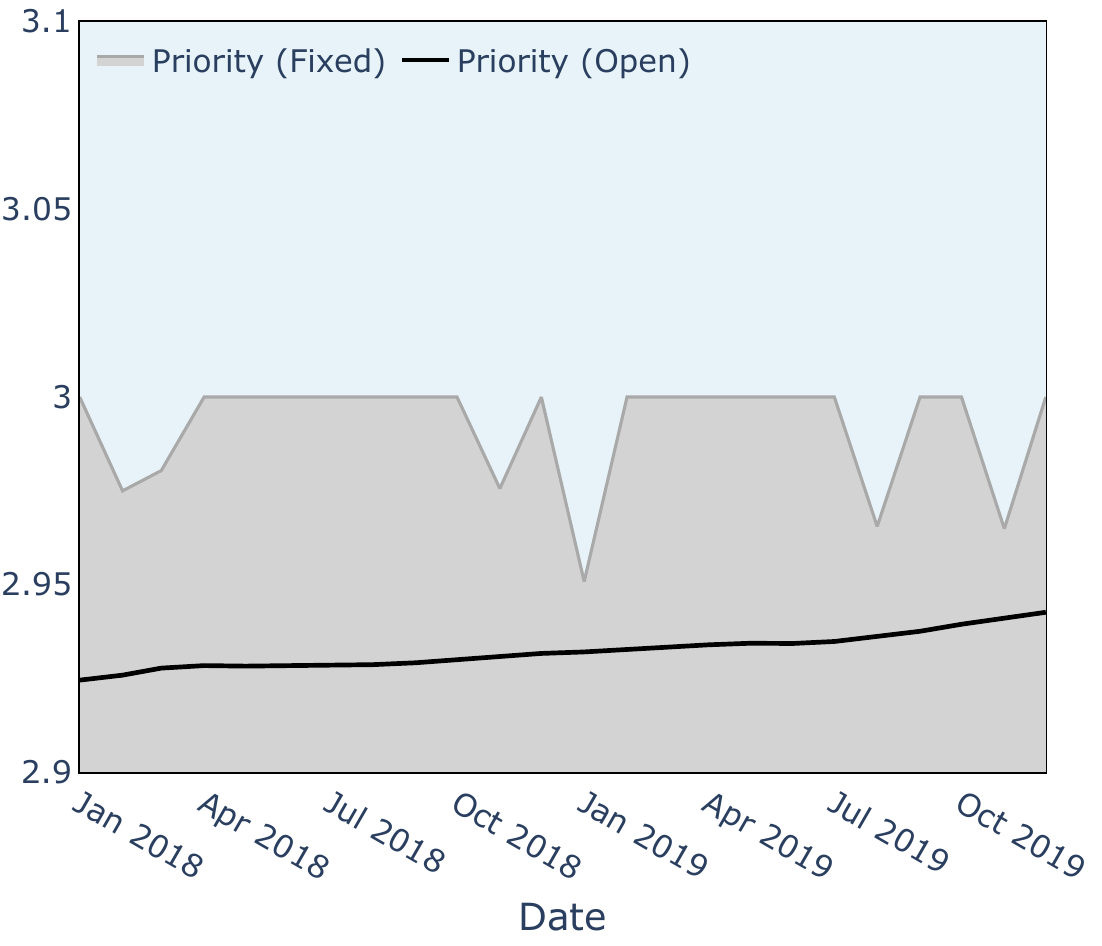}
    \caption{Eclipse (priority)} \label{fig:Eclipse_prio_solv}
  \end{subfigure} \\
  \medskip
  \begin{subfigure}[t]{.31\textwidth}
    \centering\includegraphics[width=\textwidth]{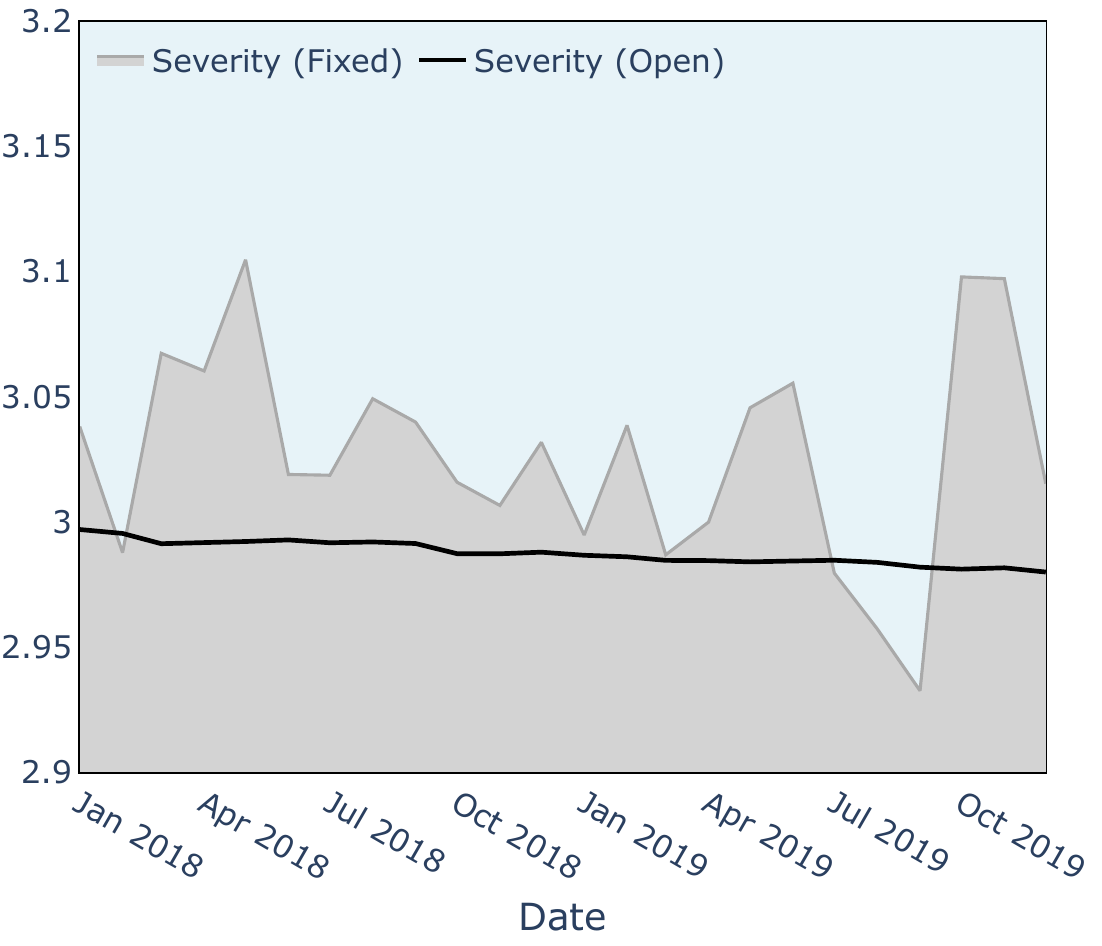}
    \caption{Mozilla (severity)} \label{fig:Mozilla_sever_solv}
  \end{subfigure}\quad
  \begin{subfigure}[t]{.31\textwidth}
    \centering\includegraphics[width=\textwidth]{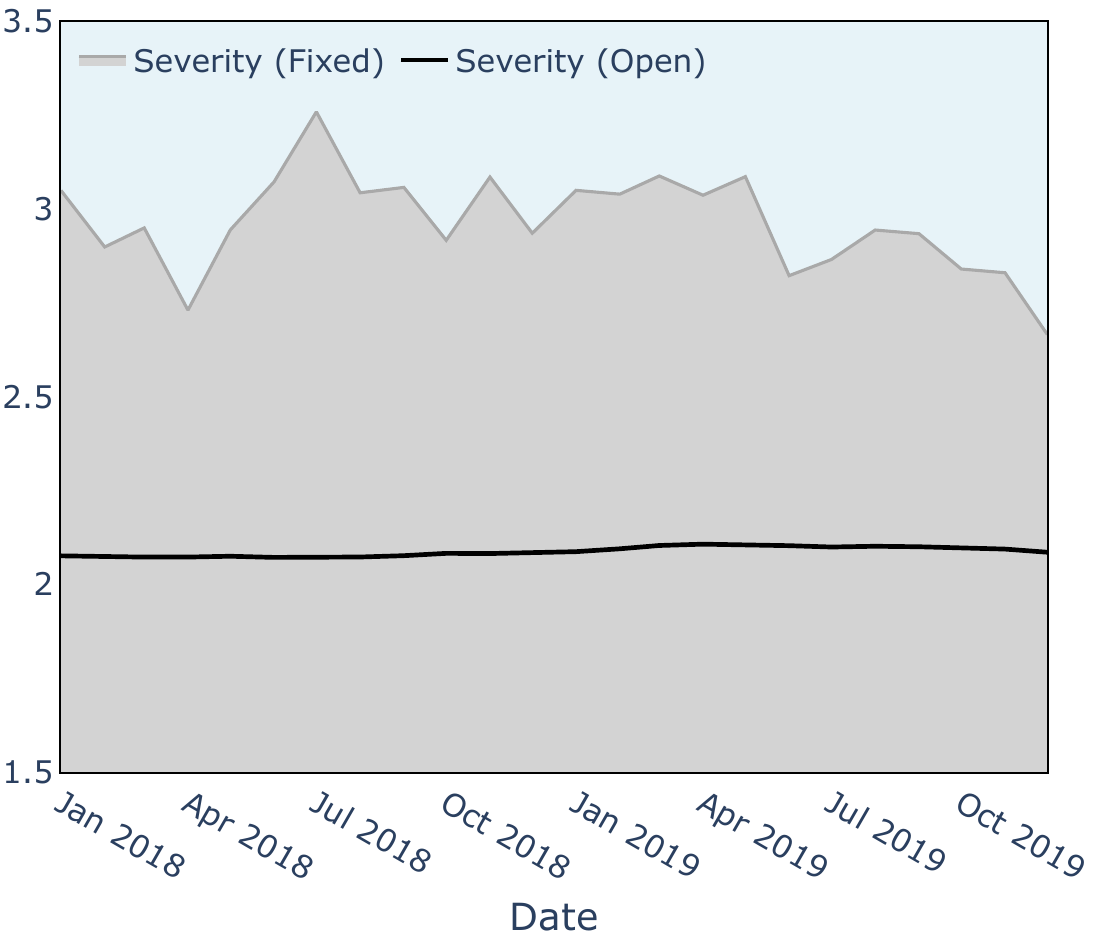}
    \caption{LibreOffice (severity)} \label{fig:LibOffice_sever_solv}
  \end{subfigure} \quad
  \begin{subfigure}[t]{.31\textwidth}
    \centering\includegraphics[width=\textwidth]{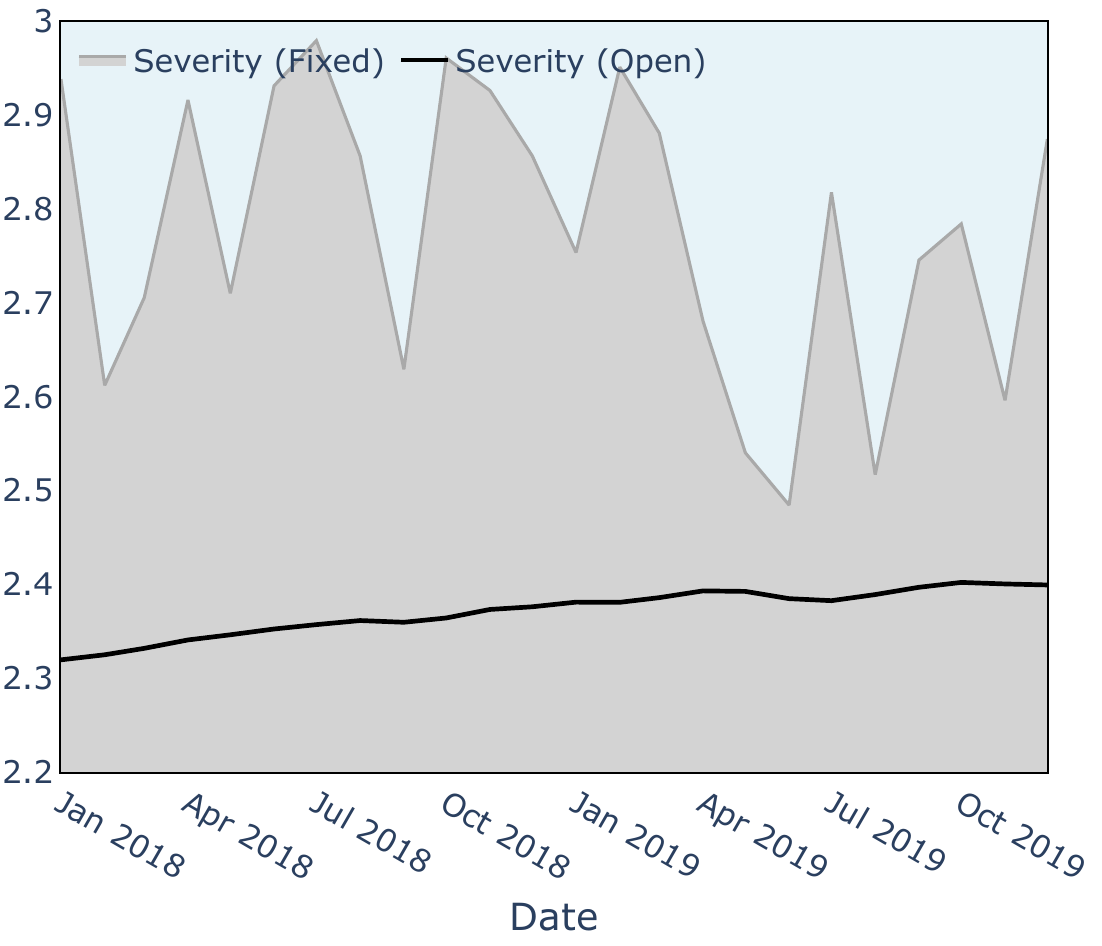}
    \caption{Eclipse (severity)} \label{fig:Eclipse_sever_solv}
  \end{subfigure}
    ~\caption{The comparison of the monthly priority and severity of the bugs in BDG and fixed bugs \textit{(the area plot shows the priority/severity of fixed bugs, whereas blue lines indicate the priority/severity of remaining bugs in the graph; $y$-axis range differs for each project.)}.}
    \label{fig:prio_sever_of_solv}
\end{figure*}

\textcolor{black}{While the subjectivity of the priority and severity can be of concern, the question of whether developers consider these subjective features in their prioritization and triage process can be answered using our proposed Wayback Machine. Specifically, we explore the evolution of the severity and priority in the ITS by comparing the mean severity and priority of the fixed bugs with those of open bugs. Figure~\ref{fig:prio_sever_of_solv} shows the average priority and severity of the fixed bugs as the grey area and the open bugs as the black line. First, we observe no significant change in the priority or severity level of the open bugs in all three projects. At the same time, we find that the average priority and severity of the fixed bugs are almost always higher than the open ones. Accordingly, we note that although these features are subjective, they are still used in practice in the triage process. On the other hand, we see that in Mozilla, the priority seems to be a more significant factor than severity, whereas, in the other projects, the reverse can be true. Referring to Table~\ref{tab:bug_info}, we emphasize many missing values for the priority level in Mozilla that we consider as the lowest level. Consequently, many of Mozilla's open bugs do not have a priority level, and the average priority level of the open bugs is close to zero. However, for the other two projects, the priority level is around three, i.e., the default value.}

\textcolor{black}{We find degree, depth, priority, and severity as important factors in the triage process; however, their significance may vary from one project to another.} To further analyze the importance of the BDG in the prioritization process, we explore the triagers' tasks in the subsequent research questions.

\subsection{Evaluating the bug prioritization and triage algorithms} \label{sec:results}

\begin{RQquestion}
    \textbf{\textcolor{black}{RQ2a: How do different bug prioritization strategies perform in terms of evolutionary metrics?}}
\end{RQquestion}

\textcolor{black}{In this research question, we investigate the prioritization module of the Wayback Machine. This module can be utilized by researchers to apply their proposed bug prioritization technique. Here we implement six different prioritization methods together with random prioritization and the actual decisions of the developers. Any other method can be incorporated into the machine and be compared with other scenarios. The Wayback Machine outputs different performance metrics, three of which are shown here, namely, the number of assigned bugs, the number of early, on-time, and late prioritization, and the standard deviation of the methods from the actual cases. Note that the second and third metrics, which we call evolutionary, can be best reported by an event regenerator that builds the exact environment at the time of prioritization.}

\textcolor{black}{We consider the assignment time of a bug as its relative importance. Specifically, we record how many times proposed prioritization strategies can assign a bug on the same day of its actual assignment. Whenever a feasible bug is assigned, we run the model to see whether it is able to prioritize the same bug over other open bugs. The same-day assignment is called ``on-time'', and the rest are defined as ``early'' or ``late''.}

\textcolor{black}{We explore the performance of different strategies on bug prioritization in the long term. The practical aim of this experiment is to see how Wayback Machine can facilitate bug prioritization performance reports in the regenerated, actual environment. We also aim to contrast the performance of different policies against the actual bug prioritization. Here, we assume that the time that a bug is assigned is its prioritization time. Therefore, we examine whether a bug prioritized by a specific method has a similar assignment/prioritization time to that of the actual prioritization. Accordingly, the assignment is considered to be a proxy for prioritization. We repeat the process for all strategies three times and report the average performance values to avoid any bias due to randomization. Table~\ref{tab:prioritization} shows the prioritization performance of different methods for different projects. ``Estimated Priority'' and ``Cost \& Priority Consideration'' have the most same-day assignment, i.e., the most similarity with the actual case. Perceived priority and fixing cost based on the textual information of the bug seems to be the most valid strategy to mimic the real cases. Interestingly, the ``estimated priority'' has much more on-time assignments than the ``maximum priority'' method. As the ML algorithm predicts the priority of a bug, it considers its relative priority given the textual information. Therefore, as the priority level is not determined for many bugs (see Table~\ref{tab:bug_info}), the model can estimate their priority levels based on the known priority. Also, the combination of the estimated priority and fixing cost considers both the important and the fast-to-resolve bugs. In that way, the strategy is able to better predict the priority of a bug. These results show the capability of Wayback Machine to objectively evaluate different prioritization strategies.}
\begin{table}[!ht]
    \caption{Summary results for different bug prioritization strategies}
    \label{tab:prioritization}
    \resizebox{\linewidth}{!}{
    \begin{tabular}{cl>{\columncolor[HTML]{EFEFEF}}rrrrrrrr}
    \toprule 
     &  & \textbf{Actual} & \multicolumn{3}{c}{\textbf{Rule-based}} & \multicolumn{3}{c}{\textbf{Machine Learning}} & \textbf{Random} \\ 
     \cmidrule(lr){4-6}\cmidrule(lr){7-9}
     &  &  & \textbf{\begin{tabular}[c]{@{}c@{}} max\\ \{depth + degree\}\end{tabular}}
     & \textbf{\begin{tabular}[c]{@{}c@{}} max\\Priority\end{tabular}} & \textbf{\begin{tabular}[c]{@{}c@{}} max\\ Severity\end{tabular}} & \textbf{Cost-oriented} & \textbf{\begin{tabular}[c]{@{}c@{}} Estimated\\ Priority\end{tabular}} & \textbf{\begin{tabular}[c]{@{}c@{}}Cost \& Priority\\ Consideration\end{tabular}} &  \\
     \midrule
    \multirow{3}{*}{{\rotatebox[origin=c]{90}{\scshape{\textbf{EclipseJDT}}}}} & \textbf{\begin{tabular}[c]{@{}l@{}}The number of \\ Assigned Bugs\end{tabular}}  & 1,251  & 1,251 & 1,251 & 1,251 & 1,251 & 1,251 & 1,251 & 1,251 \\
     \cmidrule(lr){2-10} 
     & \textbf{\begin{tabular}[c]{@{}l@{}}(Early, On-time, Late) \\ Prioritization\end{tabular}}  & (0, 1251, 0) & (810, 104, 337) & (972, 1, 278) & (821, 93, 337) & (970, 2, 279) & (349, \textbf{413}, 489) & (367, 358, 526) & (897, 21, 333) \\
     \cmidrule(lr){2-10} 
     & \textbf{\begin{tabular}[c]{@{}l@{}}Assigning Time\\ Divergence\end{tabular}} & 0 & 278 & 270 & 241 & 272 & 251 & 243 & 267 \\
     \midrule
    \multirow{3}{*}{{\rotatebox[origin=c]{90}{\scshape{\textbf{LibreOffice}}}}} & \textbf{\begin{tabular}[c]{@{}l@{}}The number of \\ Assigned Bugs\end{tabular}} & 1,570 & 1,570 & 1,570 & 1,570 & 1,570 & 1,570 & 1,570 & 1,570 \\
     \cmidrule(lr){2-10} 
     & \textbf{\begin{tabular}[c]{@{}l@{}}(Early, On-time, Late) \\ Prioritization\end{tabular}} & (0, 1570, 0) & (1188, 4, 378) & (1009, 75, 486) & (1022, 83, 465) & (1190, 1, 379) & (377, 363, 830) & (363, \textbf{370}, 837) & (1100, 331, 759) \\
     \cmidrule(lr){2-10} 
     & \textbf{\begin{tabular}[c]{@{}l@{}}Assigning Time\\ Divergence\end{tabular}} & 0 & 185 & 185 & 154 & 186 & 159 & 156 & 177 \\
     \midrule
    \multirow{5}{*}{{\rotatebox[origin=c]{90}{\scshape{\textbf{Mozilla}}}}} & \textbf{\begin{tabular}[c]{@{}l@{}}The number of \\ Assigned Bugs\end{tabular}} & 3,697 & 3,697 & 3,697 & 3,697 & 3,697 & 3,697 & 3,697 & 3,697 \\
     \cmidrule(lr){2-10} 
     & \textbf{\begin{tabular}[c]{@{}l@{}}(Early, On-time, Late) \\ Prioritization\end{tabular}} & (0, 3697, 0) & (2661, 319, 717) & (690, 764, 2243) & (3064, 59, 574) & (3162, 10, 525) & (761, 820, 2116) & (776, \textbf{861}, 2060) & (2845, 78, 774) \\
     \cmidrule(lr){2-10} 
     & \textbf{\begin{tabular}[c]{@{}l@{}}Assigning Time\\ Divergence\end{tabular}} & 0 & 126 & 162 & 122 & 123 & 146 & 143 & 135 \\
     \bottomrule
    \end{tabular}
    }
\end{table}

\begin{RQquestion}
    \textbf{\textcolor{black}{RQ2b: How do different bug triage strategies perform in terms of evolutionary metrics?}}
\end{RQquestion}

\textcolor{black}{Using the triage module of the Wayback Machine, we implement three bug triage approaches, namely, Content-Based Recommendation, CosTriage, and DeepTriage. We compare them against actual and random cases. We report six different metrics for this process to see the evolutionary performance of well-established models. We aim to investigate their average fixing time, task concentration on developers, accuracy in assigning bugs to proper developers, percentage of overdue bugs, and infeasibility of the assignments due to the blocking effect. }

\textcolor{black}{The triage process is similar to that of \citet{Kashiwa2020} and \citet{jahanshahi2021dabt}. We triage once a day and assign open bugs to available developers according to the triage algorithm. As CBR, CosTriage, and DeepTriage do not consider the available schedule of the developers, the number of assigned bugs may exceed the total capacity of a developer. Therefore, the Wayback Machine is uniquely suitable for showing the task concentration on developers since it reports both the assignment accuracy and the number of tasks assigned to each developer. In the original studies, assignment accuracy was the main concern, similar to many traditional bug triage papers. However, the Wayback Machine reveals the possibility of overdue bugs due to overwhelming experienced developers with a torrent of assigned bugs.
}

\textcolor{black}{Table~\ref{tab:bugtriage} shows the evaluation of different triage strategies based on the evolutionary metrics. To have a fair comparison, we estimate the bug fixing time for all methods using the LDA method as suggested by \citet{Kashiwa2020} and \citet{park2011costriage}. CosTriage, which considers the fixing time in its formulation, has expectedly a better average fixing time over other approaches. There is no significant difference in terms of the number of assigned developers among the three algorithms, whereas, in the case of LibreOffice and Mozilla, they assign bugs to the fewer number of developers, i.e., they overspecialize. Accordingly, they concentrate so many tasks over few top developers. The accuracy of the assignment is computed as assigning a bug to a developer who has previous experience in the same component~\cite{park2011costriage}. Using an LSTM network with attention mechanism enhances the prediction of proper developers. Since these methods concentrate tasks on a fewer number of developers, a high percentage of overdue bugs is expected. Hence, \citet{Kashiwa2020}'s work that focuses on release-aware bug triaging may address the issue. Finally, the Wayback Machine reports the infeasible assignment cases due to the blocking effect (see Table~\ref{tab:bugtriage}). This information is beneficial to the practitioners since, by definition, blocked bugs should be fixed after the blocking bugs are fixed~\citep{jahanshahi2021dabt}.}  \textcolor{black}{Our result is aligned with the conclusion of previous studies~\citep{Shirin2020, jahanshahi2021dabt} pointing to the fact that disregarding the bug dependency may incur a higher percentage of overdue bugs.}

\begin{table}[!ht]
\centering
    \renewcommand{\arraystretch}{1.3}
    \caption{\textcolor{black}{Summary results for different bug triage algorithms}}
    \label{tab:bugtriage}
    \resizebox{\linewidth}{!}{
    \begin{tabular}{cl >{\columncolor[HTML]{EFEFEF}}r rrr|r}
    \toprule
         & \textbf{} & \textbf{Actual} & \textbf{CBR} & \textbf{CosTriage} & \textbf{DeepTriage} & \textbf{Random} \\
    \midrule
        \multirow{7}{*}{{\rotatebox[origin=c]{90}{\scshape{\textbf{EclipseJDT}}}}} & \textbf{Mean Fixing Time} & \textbf{6.0} & 7.9 & 7.5 & 7.7 & 8.3 \\
        \cmidrule(lr){2-7} 
         & \textbf{The Number of Assigned Developers} & 15 & 19 & 19 & 19 & 21 \\
         \cmidrule(lr){2-7} 
         & \textbf{Task Concentration $(\mu \pm\sigma)$} & $83.4\pm93.7$ & $65.8\pm112.0$ & $65.8\pm108.5$ & $72.1\pm102.2$ & $57.5\pm88.3$ \\
         \cmidrule(lr){2-7} 
         & \textbf{Assignment Accuracy} & \textbf{97.7} & 95.5 & 94.0 & 96.7 & 38.1 \\
         \cmidrule(lr){2-7} 
         & \textbf{Percentage of Overdue Bugs} & \textbf{66.0} & 82.2 & 79.6 & 78.3 & 89.3 \\
        \cmidrule(lr){2-7} 
         & \textbf{Infeasible Assignment w.r.t. the BDG} & \textbf{5.4} & 6.0 & 5.8 & 6.3 & 5.9 \\
        \midrule
        \multirow{7}{*}{{\rotatebox[origin=c]{90}{\scshape{\textbf{LibreOffice}}}}} & \textbf{Mean Fixing Time} & 3.3 & 2.1 & \textbf{1.8} & 1.9 & 2.3 \\
        \cmidrule(lr){2-7} 
         & \textbf{The Number of Assigned Developers} & 57 & 22 & 21 & 23 & 23 \\
         \cmidrule(lr){2-7} 
         & \textbf{Task Concentration $(\mu\pm\sigma)$} & $27.5\pm68.9$ & $71.3\pm224.5$ & $74.7\pm253.2$ & $70.7\pm218.4$ & $66.1\pm173.7$\\
         \cmidrule(lr){2-7} 
         & \textbf{Assignment Accuracy} & 91.7 & 99.1 & 99.3 & \textbf{99.4} & 43.3 \\
         \cmidrule(lr){2-7} 
         & \textbf{Percentage of Overdue Bugs} & \textbf{35.9} & 77.1 & 80.8 & 76.2 & 81.3\\
         \cmidrule(lr){2-7} 
         & \textbf{Infeasible Assignment w.r.t. the BDG} & \textbf{0.1} & 0.1 & 0.1 & 0.1 & 0.2 \\
        \midrule
        \multirow{7}{*}{{\rotatebox[origin=c]{90}{\scshape{\textbf{Mozilla}}}}} & \textbf{Mean Fixing Time} & 7.0 & 7.2 & \textbf{6.6} & 7.1 & 8.6 \\
        \cmidrule(lr){2-7} 
         & \textbf{The Number of Assigned Developers} & 137 & 74 & 85 & 80 & 115  \\
         \cmidrule(lr){2-7} 
         & \textbf{Task Concentration $(\mu\pm\sigma)$} & $27.0\pm49.5$ & $50.1\pm204.0$ & $43.6\pm187.0$ & $41.7\pm192.3$ & $31.5\pm42.3$ \\
         \cmidrule(lr){2-7} 
         & \textbf{Assignment Accuracy} & \textbf{72.7} & 60.2 & 59.0 & 62.1 & 15.5 \\
         \cmidrule(lr){2-7} 
         & \textbf{Percentage of Overdue Bugs} & \textbf{69.8} & 80.1 & 77.6 & 78.5 & 82.6 \\
         \cmidrule(lr){2-7} 
         & \textbf{Infeasible Assignment w.r.t. the BDG} & 9.4 & 9.0 & \textbf{8.8} & 9.8 & 11.2 \\
    \bottomrule
    \end{tabular}
    }
\end{table}

\textcolor{black}{Without considering the evolutionary nature of the reported bugs in the ITS, reporting accuracy of the bug triage model might be misleading. Therefore, the Wayback Machine provides a tool for researchers to explore other impacts that their proposed model may have on the whole ecosystem.}

\subsection{\textcolor{black}{Wayback Machine validation}}
We check the validity of our Wayback Machine in \textcolor{black}{four} main steps.
First, we compare the experiment outputs with the reported results in the previous studies.
We observe that the main outcomes for the considered prioritization/triage algorithms largely overlap with those of the previous works.
For instance, \citet{Kashiwa2020} used a simple simulator and found a reduction in the fixing time of CosTriage compared to that of CBR (between 10\% and 33\% improvement). 
Our Wayback Machine, using different extracted datasets, also demonstrates a better performance for CosTriage compared to CBR in terms of the average fixing time (between 6\% and 16\% improvement). 
We juxtapose the results obtained by the Wayback Machine with that of \citet{Kashiwa2020} in terms of the number of assigned developers, accuracy, and task distribution. 
We observe that our findings are consistent with the reported results, pointing to the validity of the Wayback Machine.
We also compare certain outcomes (e.g., number of bugs over time) with the raw extracted data to see whether various event counters in the simulation work as intended.

Second, we conduct a sensitivity analysis over particular model parameters to understand whether the Wayback Machine demonstrates the expected behavior based on the parameter values. 
For instance, we observe how the percentage of overdue bugs changes with the number of developers; that is, we impose different constraints on the total number of available developers to see how the number of unfixed/overdue bugs changes as we add more developers to the system. 
As expected, the results indicate an inverse correlation between the number of overdue bugs and developers.
Such analyses indicate that the Wayback Machine and its associated data are an accurate representation of the underlying ITS. 
We conduct various sensitivity analyses, which help us fix potential issues in the Wayback Machine implementations.

\begin{figure}[!ht]
    \centering
    \subfloat[Mozilla (CSV) \label{fig:MozillaCSV}]{\includegraphics[width=0.33\textwidth]{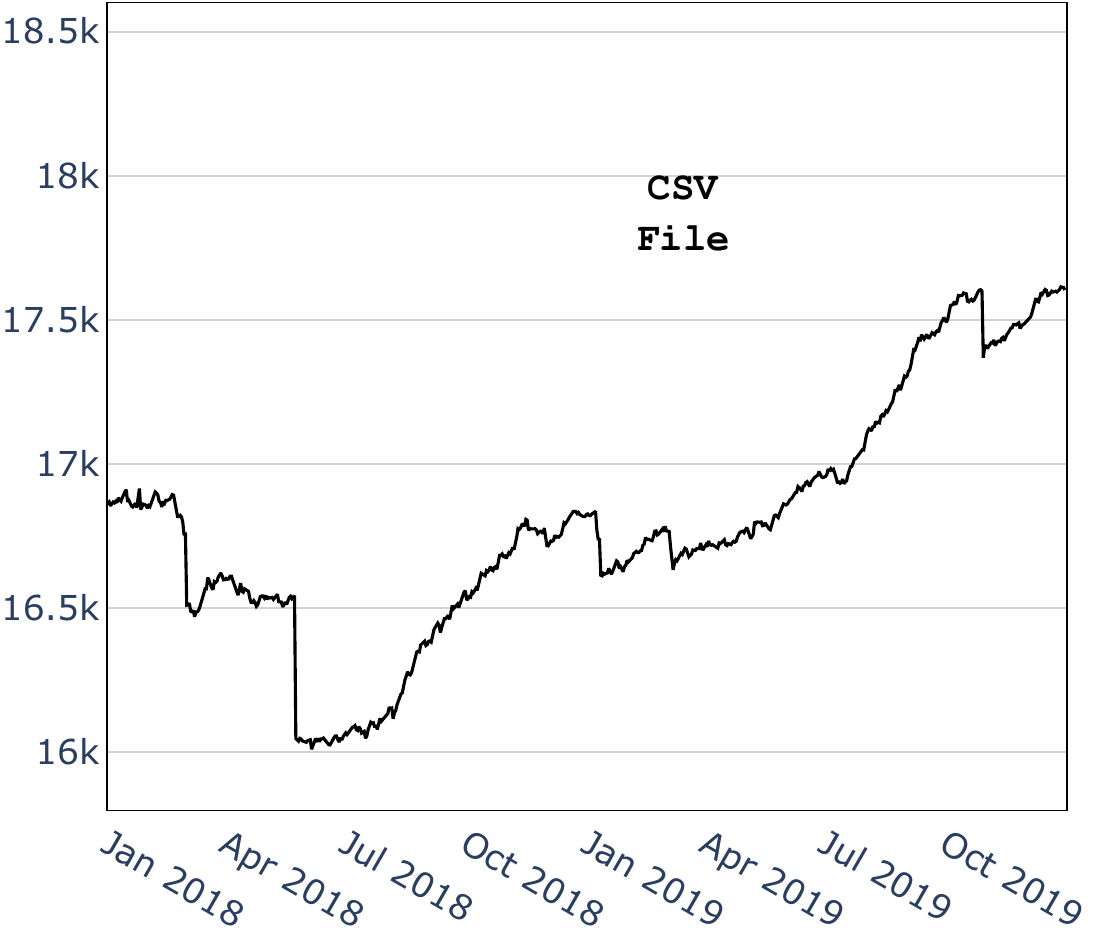}}
    \hfill
    \subfloat[LibreOffice (CSV) \label{fig:LibreOfficeCSV}]{\includegraphics[width=0.33\textwidth]{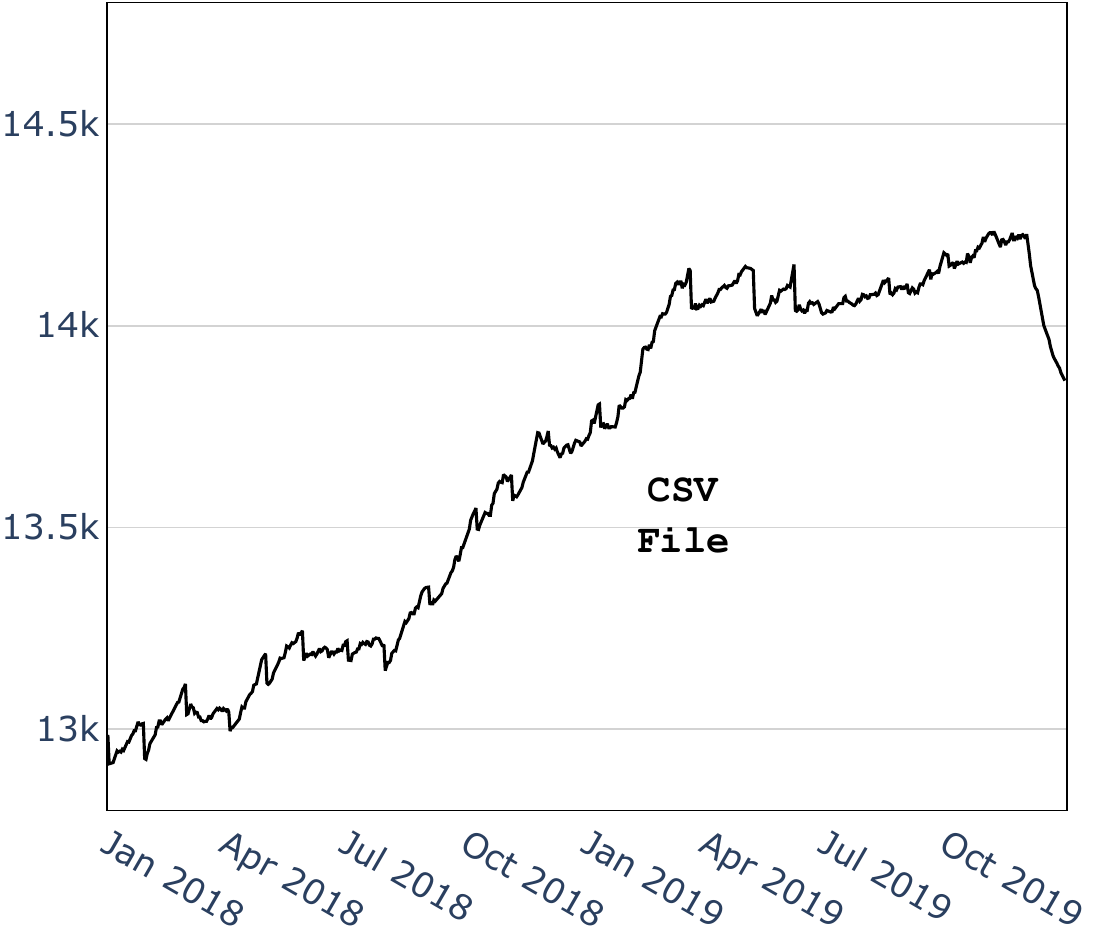}}
    \hfill
    \subfloat[Eclipse (CSV) \label{fig:EclipseCSV}]{\includegraphics[width=0.33\textwidth]{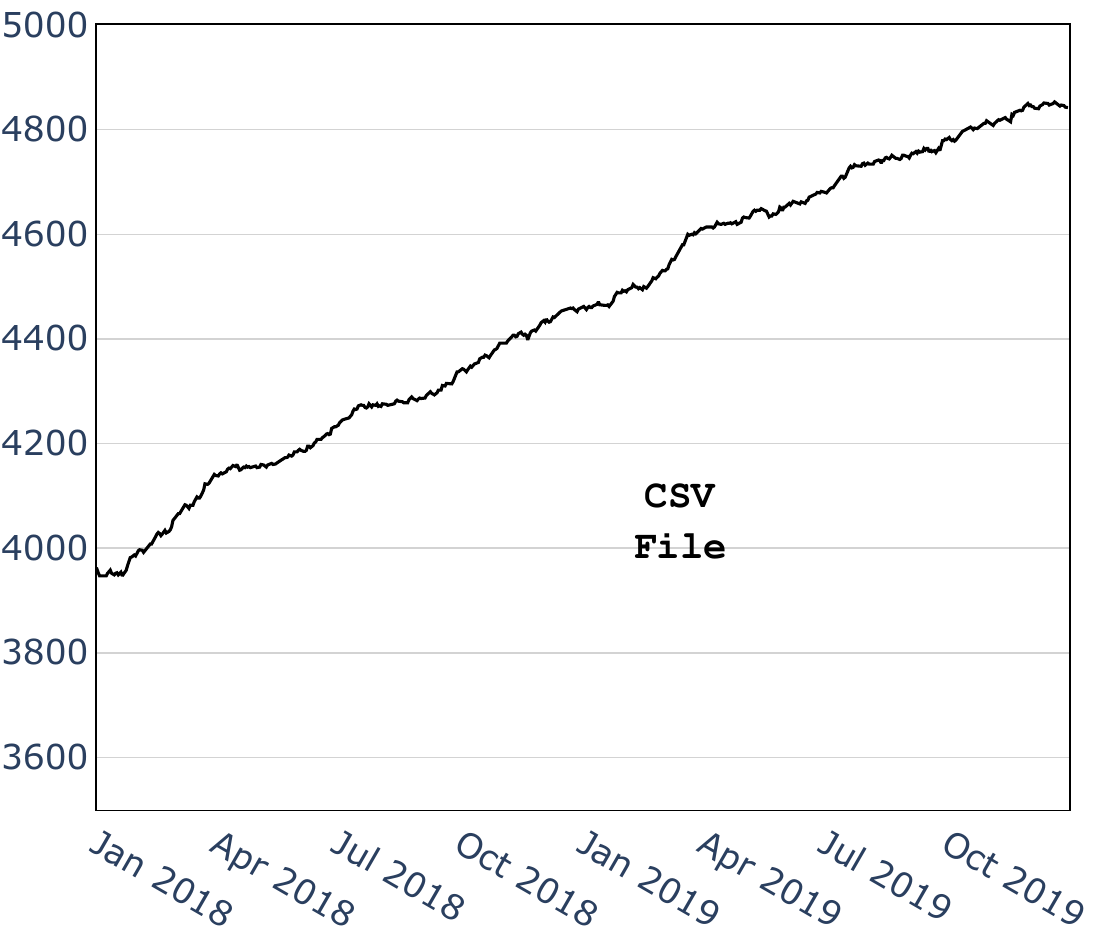}}
    \\
    \subfloat[Mozilla (Wayback Machine) \label{fig:MozillaWayback}]{\includegraphics[width=0.33\textwidth]{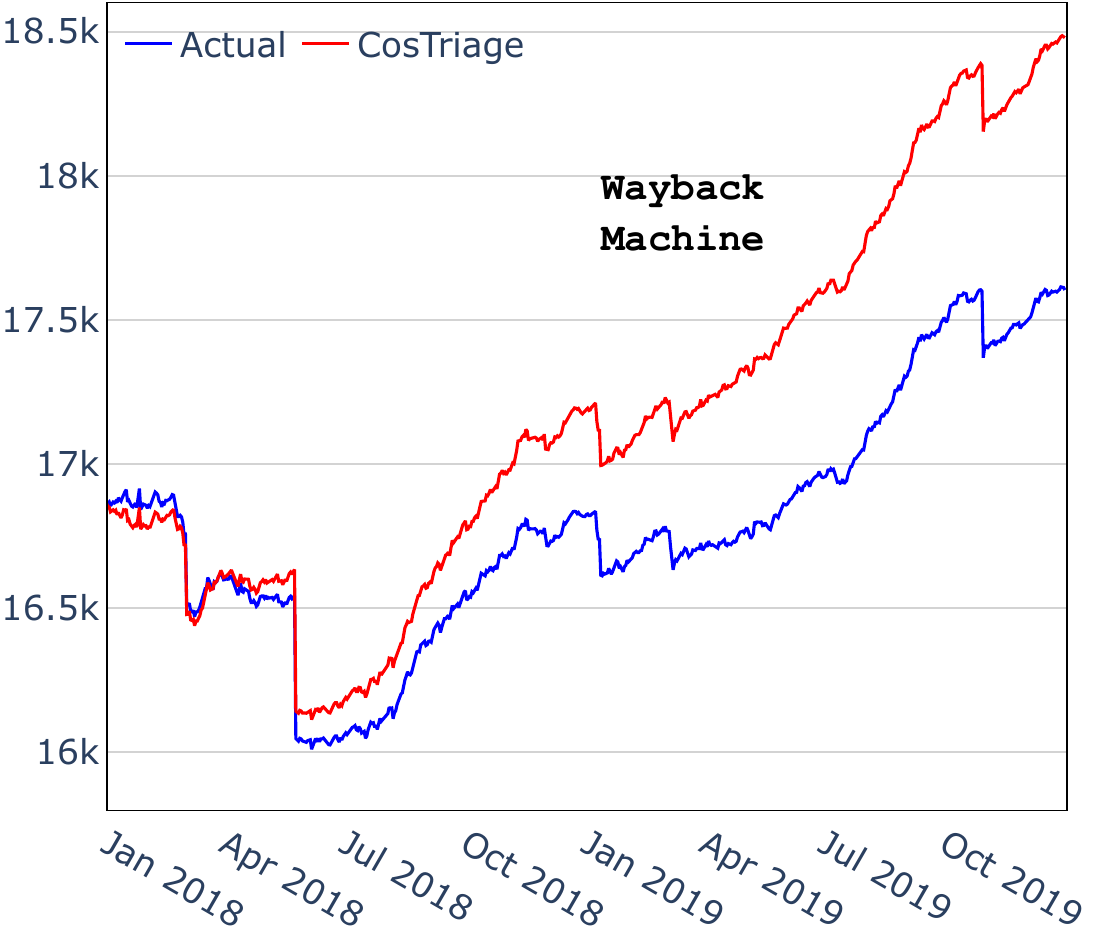}}
    \hfill
    \subfloat[LibreOffice (Wayback Machine) \label{fig:LibreOfficeWayback}]{\includegraphics[width=0.33\textwidth]{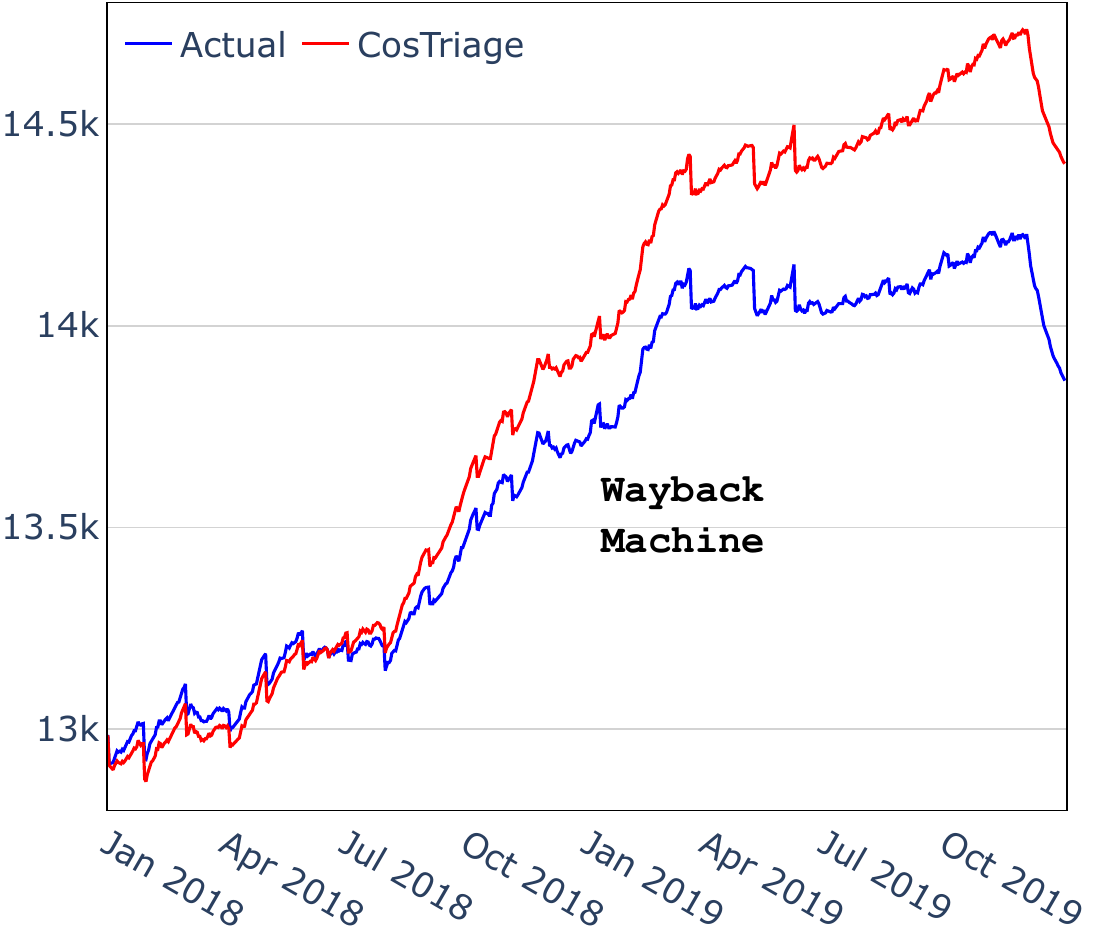}}
    \hfill
    \subfloat[Eclipse (Wayback Machine) \label{fig:EclipseWayback}]{\includegraphics[width=0.33\textwidth]{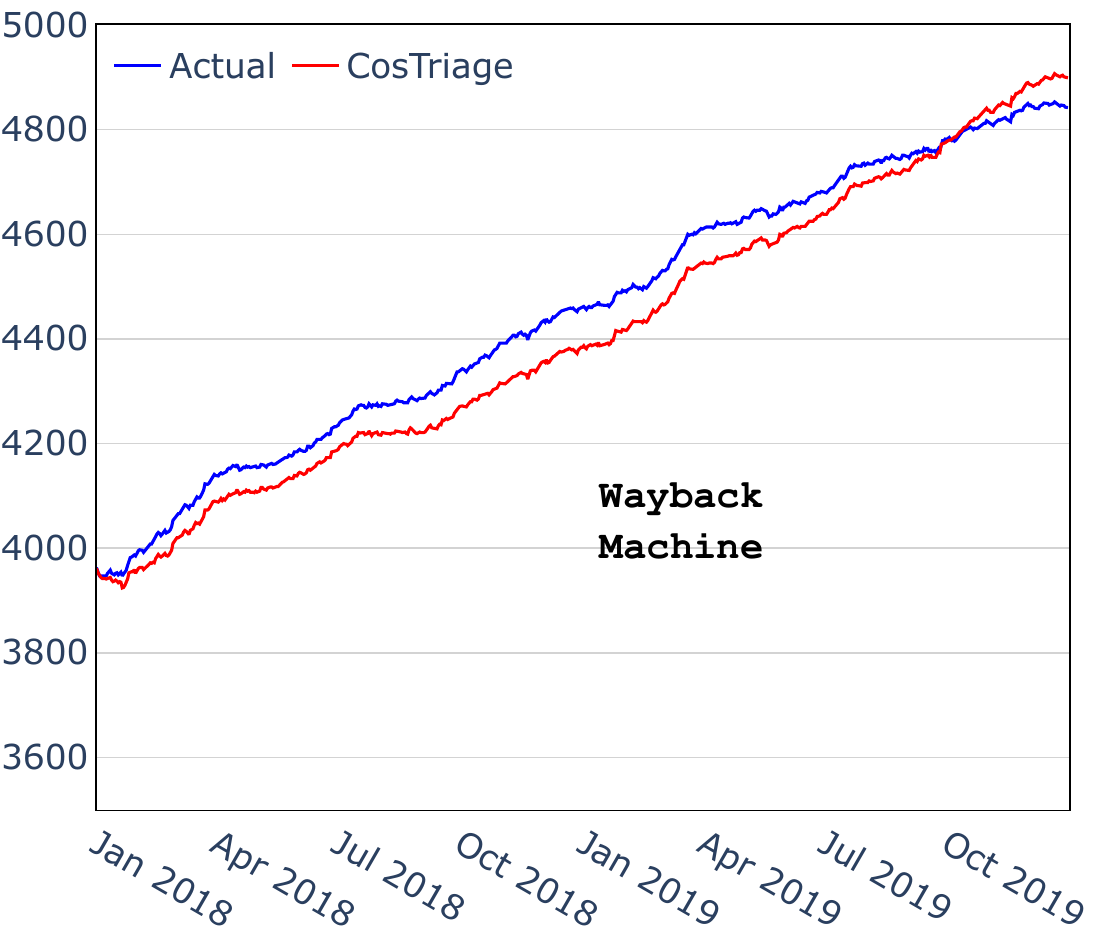}}
    \caption{\textcolor{black}{The number of unfixed bugs in the ITS obtained through the static approach and past-event regenerator approach.}}
    \label{fig:number_of_bugs}
\end{figure}

\textcolor{black}{Third, we evaluate the Wayback Machine by comparing its output with the statistics directly obtained from the ITS, which we extract as CSV files.
Since the Wayback Machine contains a large number of counters to track the events, such a comparison would serve as a sanity check for those counters. 
For instance, Figure~\ref{fig:number_of_bugs} shows the number of bugs accumulated in the ITS as reported by the Wayback Machine and in the CSV file. 
We observe a consistent result with that of the CSV file. 
On the other hand, it is important to note that the traditional use of CSV files for bug triage automation fails to address bug accumulation in the system while using a specific prioritization/triage method, and the Wayback Machine is particularly suited for such purposes. 
Bug accumulation corresponds to the gradual increase of the number of bugs in the system, creating a ``bug backlog''. 
Such a case is not easily manageable by the triagers/developers, and leads to overdue bugs.
\citet{park2016cost} referred to the bug accumulation issue as overspecialization due to continually recommending the bugs to the most-experienced developer, which results in overloading of this developer. 
In other words, when a model assigns too many bugs to a few developers, their schedules could be overloaded to the point that many unresolved bugs become accumulated in the system.
In Figure~\ref{fig:number_of_bugs}, we also illustrate the outcomes for a specific method (e.g., CosTriage), which serves as a use case for the Wayback Machine to assess the effectiveness of a prioritization/triage method.}

\textcolor{black}{
We note that some metrics, such as bug assignment accuracy, can be reported without having any past-event regenerator (i.e., Wayback Machine). 
Any ML model, e.g., SVM, can be trained and predict the assigned developers. 
Therefore, we applied SVM once within Wayback Machine, which predicts assignments one by one on a daily basis, and once without Wayback Machine and using the (static) CSV file of the bug attributes. 
The accuracy values obtained by these two approaches are found to be identical.
We consider this observation as another sanity check for the soundness of the Wayback Machine.
}

\begin{figure}[!ht]
    \centering
    \subfloat[Mozilla \label{fig:fixing_datesMozilla}]{\includegraphics[width=0.33\textwidth]{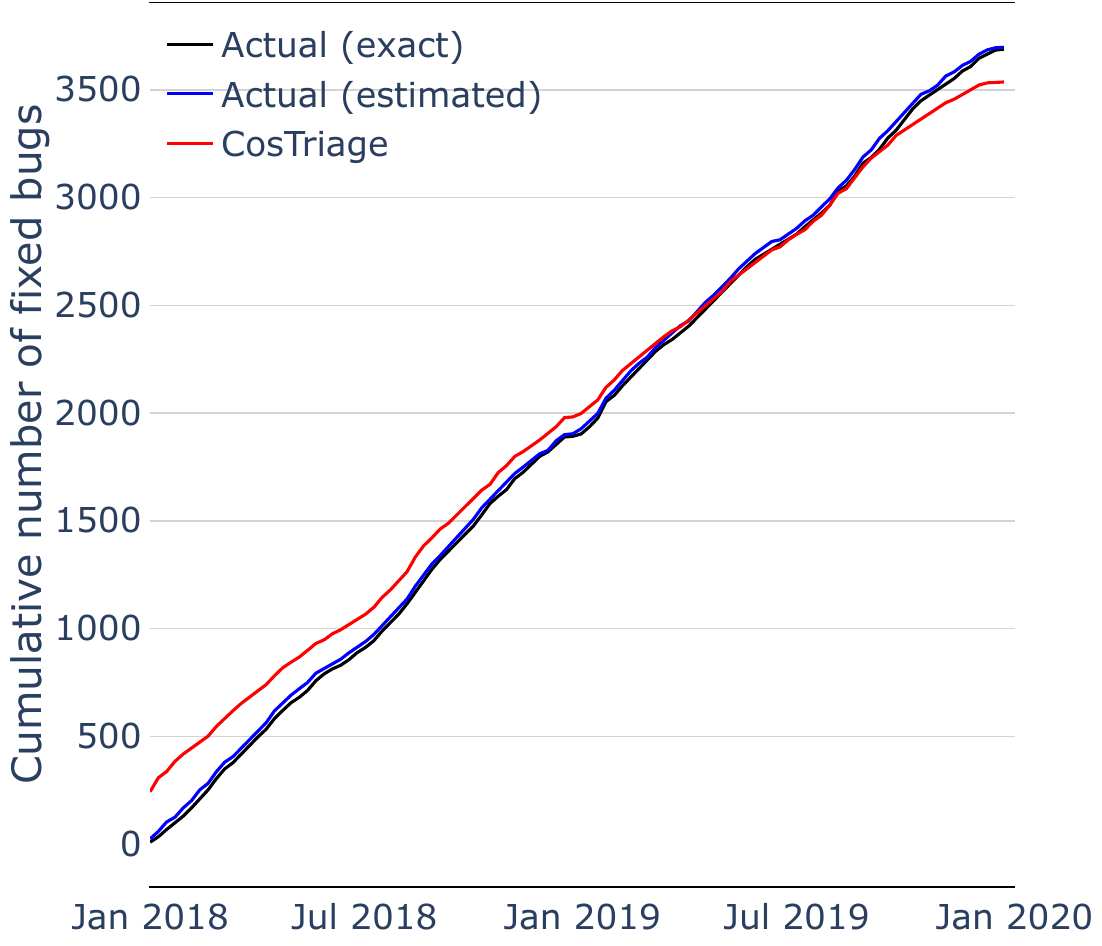}}
    \hfill
    \subfloat[LibreOffice \label{fig:fixing_datesLibreOffice}]{\includegraphics[width=0.33\textwidth]{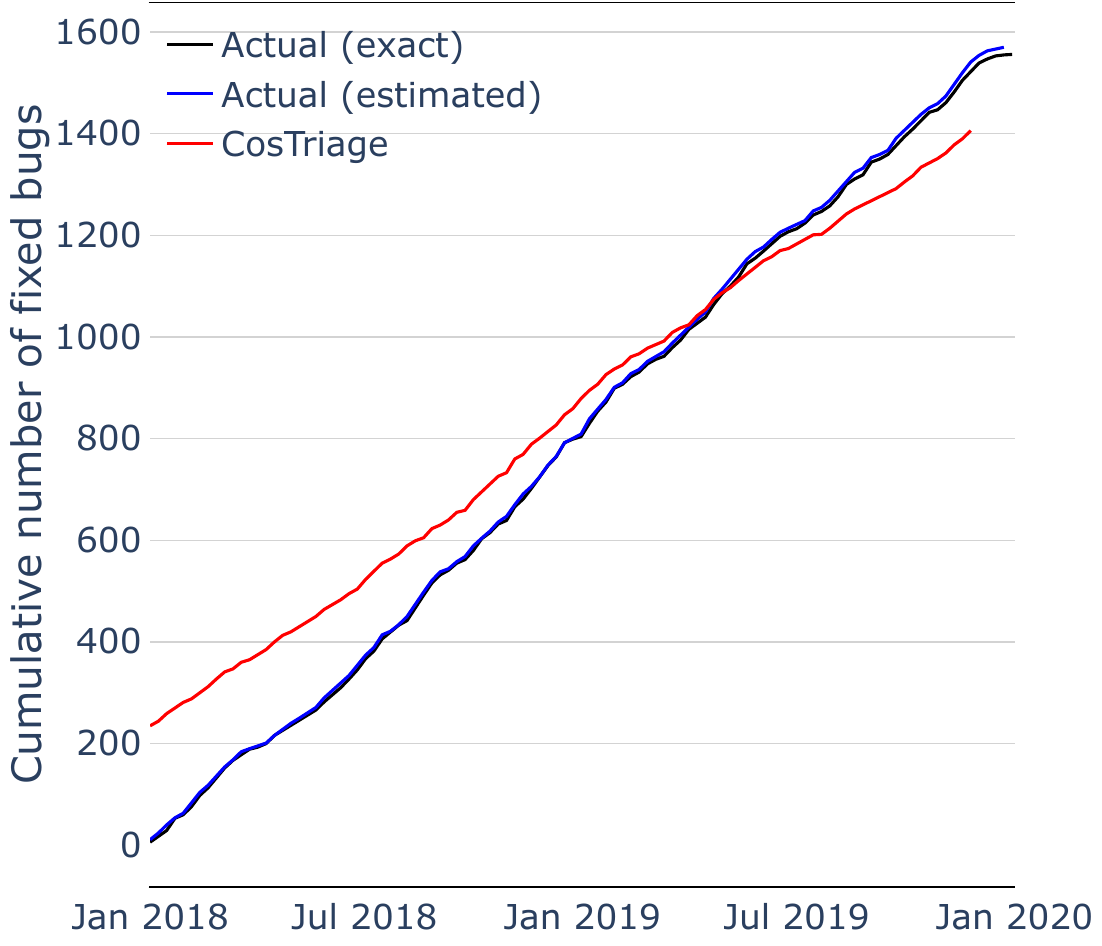}}
    \hfill
    \subfloat[Eclipse
    \label{fig:fixing_datesEclipse}]{\includegraphics[width=0.33\textwidth]{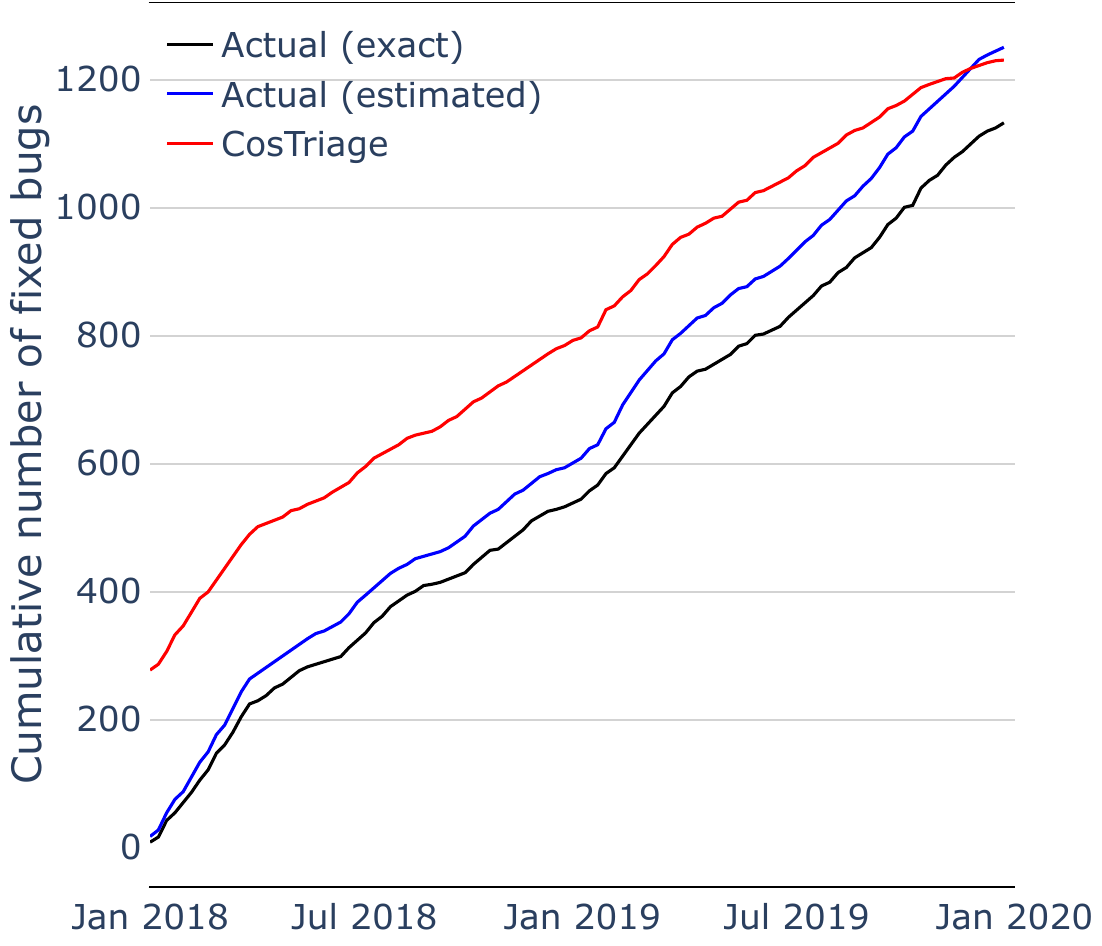}}
    \caption{\textcolor{black}{The cumulative number of bugs fixed using exact dates (CSV file), estimated fixing dates (Wayback Machine), and CosTriage.}}
    \label{fig:fixing_dates}
\end{figure}

\textcolor{black}{
As a multi-step validity check of the model, we compared the fixing date of the bugs using the actual fixing date (i.e., derived from the CSV file), estimated fixing date (i.e., derived from the LDA and collaborative filtering~\cite{park2011costriage} in the Wayback Machine), and CosTriage's fixing date. 
Figure~\ref{fig:fixing_dates} shows the cumulative number of bugs that are fixed during the testing phase. 
In this exercise, a bug can be assigned either to the ground-truth developer (i.e., by using the actual information in the ITS) or using CosTriage to any (other) suitable developer. 
CosTriage does not have a built-in schedule of the developers and cannot decide on the exact date of assignment. 
Therefore, the figure shows a clear difference between that model and the actual case in terms of the cumulative number of fixed bugs. 
However, whether we use the estimated fixing time for a bug or its actual fixing time, there is a distinct overlap between the Wayback Machine and the actual case. 
This observation indicates that the minor discrepancy is due to employing the LDA within the machine and not the implementation, pointing to the validity of the Wayback Machine. 
This conclusion is aligned with Table 6 in \citet{Kashiwa2020}'s paper, in which they showed the difference between the simulated and actual fixing days.
}

\textcolor{black}{
Lastly, certain summary statistics cannot be obtained without having a simulator. 
For instance, the percentage of overdue bugs is time-dependent, and whether a bug is overdue or not depends on other bugs assigned to a developer and the current schedule of the same developer. 
Such delicate analysis indicates the importance of having an event-regenerator that gives a new dimension (e.g., time) to bug triage analysis.
Hence, in the case of time-dependent metrics, we do not have a baseline to compare with, and we run the algorithm in ``Run and Debug'' mode and check every step meticulously to ensure there are no bugs/errors in our implementations. 
} We further validate our model based on expert opinions by sharing our findings with the developers from the corresponding ITS. 
They comment on our work regarding possible data, implementation, and interpretation issues, which serve as another important sanity check.

\section{Threats to validity} \label{sec:threats}
We summarize the threats to the validity of our study as follows.

\paragraph{Construct validity}
We report the model performance based on the train-test split, where the train set consists of the data from 2010 to 2018, and the test set period is taken as 2018 and 2019. On the other hand, the ITS is evolving, and some definitions may change while we split the data in this way. For instance, some active developers in 2012 may become inactive in 2019 and leave the system. Moreover, introducing new features for the software produces new bugs that do not exist in history. We disregard developers that are inactive for the past two years or whose activities have been reduced significantly. Additionally, a rolling train-test split strategy can alleviate this issue. However, we rely on the common practice and definitions from previous studies, and we take a similar approach for all strategies to make them comparable~\cite{Kashiwa2020}. Furthermore, we consider the changes in all attributes during the life-cycle of a bug. For instance, whenever a dependency is found, we add it to the bug attribute, and we do not use the last status of the bugs. Nevertheless, the changes in the severity level are not directly extractable from bug history. Therefore, we leave exploring how the changes in perception of severity impact the bug prioritization and triage outcomes to future research. 

In this study, we regenerated past events in the ITS of three projects. We further applied different prioritization and triage algorithms used in the literature. We made certain assumptions for each case, which are consistent with the previous studies. 
However, we acknowledge that those assumptions might be strong, and bug prioritization/triage as a multifaceted problem may not be handled by simple, naive approaches in practical settings.
Moreover, we evaluated different strategies in terms of evolutionary metrics, e.g., the number of overdue bugs, together with traditional metrics, e.g., the assignment accuracy. That is, we ensure to include a complete list of metrics that can be used by other researchers while reporting their model's performance. Nevertheless, the Wayback Machine can easily incorporate more metrics based on the study objectives. Regarding the evolution in severity levels of a bug, textual data of each bug's discussions needs to be mined to extract necessary information as the changes in the severity level are not directly recorded in the bug's history. We plan to extend our work by incorporating the dynamics of the severity levels during a bug's lifespan.

\paragraph{External validity}
In our experiments, we rely on the data extracted from three different open-source projects, namely, Firefox, Eclipse, and LibreOffice.
Specifically, we choose well-established projects with different characteristics, covering the associated bug information for the past decade. \textcolor{black}{Nonetheless, replication of our study for industrial data or proprietary products would prove fruitful.} 
We also consider the evolution of the bug reports instead of static snapshots of the system. We simplify our models by discarding some attributes, e.g., the number of CC'ed developers or comments' contents. We plan to expand our study by including different attributes of bug reports and create a more comprehensive evolutionary machine. We used the actual bug prioritization obtained from the ITS as the baseline, and since, to the best of our knowledge, there is no other study that reproduced bug prioritization or triage events, we incorporated other works according to our defined mechanism.
\textcolor{black}{As some strategies in our experiment have randomness in their process, i.e., they randomly choose a bug in the case of ties, we repeat all experiments multiple times and report the results based on their average performance. We expect this approach to address the issue of random heterogeneity of subjects.}

\paragraph{Internal validity}
The BDG is extracted from three Bugzilla ITS using the REST API. However, some bug reports might be deleted from the repository or have limited access to normal users. Our analysis applies to the bugs that are open to the public. \textcolor{black}{Furthermore, we estimate fixing time using the formulation proposed by~\citet{park2011costriage}, that is, ``$\textit{fixing time} = \textit{fixing date} - \textit{assignment date} + 1$''. Nevertheless, we acknowledge that the exact solving time for a bug cannot be determined beforehand.} Therefore, all reported fixing times in the result section are estimated times to solve bugs. This assumption is not considered to impact the final decision when comparing different strategies since it remains identical for each strategy.

\section{Related work} \label{sec:background}
Bug prioritization \textcolor{black}{and triage are} vital tasks in software systems as they affect the maintenance budget of software, scheduled releases and enhancements, and even the image of a brand in the eyes of end-users. The developers typically use manual examination and intuitive judgment in the process of bug triage. \citet{valdivia2016} reports that there is no specific bug prioritization strategy on which developers agree during the bug fixing process.

Bug triaging involves different processes such as designating an appropriate developer with relevant knowledge to resolve a bug, analyzing the time to fix a bug, specifying which bug needs to be solved immediately and which one does not, and finding duplicate bug reports~\citep{uddin2017}. Therefore, manual implementation of such an arduous process requires considerable time and resources in large and open-source software systems, making this task error-prone. A considerable amount of research aims to alleviate this issue through the automation of the entire triaging process. For instance, researchers approach the problem of duplicate bug detection using text retrieval techniques or more complex learning-based methods, including additional bug information~\citep{Chaparro2019, hindle2016, EBRAHIMI2019, hindle2019}. On the other hand, several other studies focused on automatic or semi-automatic bug triage models to either select the bug which should be solved next or choose an appropriate developer to solve it~\citep{jahanshahi2021dabt, Umer2018,Zhang2017,guo2020}.

In terms of bug triaging, different machine learning approaches, such as classification, integer programming, information retrieval, and reinforcement learning were adopted. \textcolor{black}{\citet{park2011costriage}, referring to the over-specialization of content-based recommendation (CBR), considered both accuracy and fixing cost in their formulation. They combined CBR with a collaborative filtering recommender (CBCF). They use the Latent Dirichlet Allocation (LDA) approach to enhance the quality of the CBCF method.} \citet{Yang2014} suggested a method for semi-automatic bug triage and severity prediction. They utilized topic modeling, e.g., LDA, to determine the topic to which an arriving bug belongs. Then, they extracted a list of candidate assignees based on the selected topic and used bug attributes to rank appropriate developers. Similarly, \citet{Xia2017} proposed multi-feature topic model, which is an extensible topic model based on the LDA approach that computes the affinity of a developer to a new bug report using the history of the bugs the developer has ever fixed. \textcolor{black}{\citet{Kashiwa2020} used an integer programming (IP) formulation to address overdue bugs. They also improved the previous works by setting a limit on developers' capacity to solve bugs simultaneously. \citet{jahanshahi2021dabt} used our proposed Wayback Machine and improved \citet{Kashiwa2020}'s work by adding a constraint on bug dependency. They further reduced the fixing time by changing the IP objective function and embedding the fixing cost there. In this regard, our contribution to the literature includes a past-event regenerator facilitating performance report of the triage models, incorporating some important methods from the literature along with evolutionary performance metrics, and comparing their results with the actual sequence of historic decisions. }

Regarding bug prioritization, \citet{Umer2018} studied the effect of emotion analysis for the summary attribute of bug reports on bug prioritization. Specifically, they computed the emotion-value of each bug report and assigned them a priority level of P1 to P5. Moreover, they reported a high correlation ($r=0.405$) between emotion and the priority of bug reports. \citet{guo2020} utilized natural language processing methods using Word2vec representation of bug summary and implemented a convolutional neural network (CNN) to determine bug triaging decisions. \citet{Shirin2020} pointed to a different concern for bug prioritization, noting that the bug priority and severity can be both subjective and misleading. They focused on the mutual impact of bugs by using a dependency graph. Although few other studies consider a graph-based analysis for the software evolution~\citep{Bhattacharya2012}, \citet{Shirin2020}'s work differs from those in terms of incorporating the uncertainty in the ITS. More specifically, they proposed a partially observable bug dependency graph, where the dependencies between the bugs are not fully observable beforehand and are revealed as the bugs are resolved, and defined its depth and degree as crucial factors affecting a bug's priority. They solved their POMDP model using the Monte Carlo simulation and compared the performance of the resulting policy against the baseline policies. On the other hand, their work lacks an internal performance index that would allow them to compare different policies. \textcolor{black}{In this regard, our contribution to the bug prioritization literature includes suggesting a comprehensive list of evolutionary and traditional metrics for reporting the performance of any prioritization or triage algorithm. We also consider a list of rule-based and machine learning strategies to cover different bug prioritization policies. Moreover, the novel Wayback Machine enables practitioners to compare their suggested approaches with the actual practice recorded in the ITS. Unlike previous works, we consider evaluating prioritization and triage algorithms through reconstructing the exact ecosystem at the time a decision is made. Therefore, instead of extracting bug attributes and using a fixed dataset typically stored in a CSV file to estimate bugs' priority level or the assigned developer, we rely on an evolving system that considers the exact bug attributes at each timestamp and shows the real impact of the prioritization or triage decisions.}

\section{Conclusion} \label{sec:conclusion}
\textcolor{black}{In this work, we design a Wayback Machine that regenerates past events related to bug reports in ITSs while considering the BDG, which is considered to be a reliable information source for bug prioritization and triage tasks~\cite{Shirin2020, jahanshahi2021dabt, Bhattacharya2012-2}. A detailed implementation of the Wayback Machine requires tackling three challenges. First, it needs to consider different elements of the ITS, such as users, bugs, developers, and the BDG. Second, it should be designed in a modular format to facilitate adopting any prioritization and triage algorithm. Accordingly, it can be utilized by other researchers to have a complete performance report of their prioritization and triage approaches. Most importantly, the past-event regenerator (i.e., the Wayback Machine) should comprehensively reproduce past prioritization/triage decisions and provide insight into their impacts on different system components.}


\textcolor{black}{Our work on open-source data indicates the importance of using a history regenerator that is able to implement proposed bug prioritization and triage algorithms, considering the whole ecosystem rather than applying them in a vacuum. We first explore the history of the events and the evolutionary characteristics of the bugs, e.g., severity, priority, depth, and degree. We compare the features of the resolved bugs with those remaining open during the same period. Our observations reveal the importance of bug dependency in projects with well-reported blocking effects. Moreover, we find that priority and severity, although subjective, are still significant factors in the triage process. } 

\textcolor{black}{We extend our past-event regenerator(Wayback Machine) to a mechanism that is able to integrate any bug prioritization or triage model. We embed some bug prioritization (e.g., rule-based and machine learning algorithms) and bug triage algorithms (e.g., CBR, CosTriage, and DeepTriage) into the Wayback Machine. Currently, the model tracks the algorithms' performance using evolutionary and traditional metrics through their life cycle. The machine requires bugs' information and history together with developers' information as inputs and produces detailed analysis for the given training and testing phase. Researchers may employ the Wayback Machine to have an easy-to-use evaluation tool for reporting the performances of their proposed models.}

\textcolor{black}{To validate the Wayback Machine, we utilize the data extracted from three OSS systems in Bugzilla. Our prioritization and triage experiments demonstrate novel perspectives towards the performance of the model. For instance, we observe that most models ignore the bug dependency during their triage phase. Moreover, the models overspecialize and assign tasks to few highly experienced developers. In that case, they increase their accuracy by ignoring the fact that the high number of reported bugs to the ITS requires an extensive list of developers to address them. Thus, we further explore the fairness of the task distribution and its impact on the overdue bugs. These findings were not easily achievable without the help of regenerating the exact ecosystem at the decision time.}

Our primary objective in this longitudinal study is to demonstrate the current status of the system and sequential decisions of the developers in these projects to facilitate exploring different bug prioritization \textcolor{black}{and triage} strategies. 
\textcolor{black}{For practitioners, it highlights the importance of the history of the ITS in bug prioritization and triage. It also facilitates the comparison of any strategy with the actual decision-making process.} In the end, we recommend considering the evolutionary behavior of the ITS instead of snapshots of the past events, and a simulation study similar to ours would be helpful for this purpose.

\section*{Supporting information}

To make the work reproducible, we publicly share our originally extracted dataset of one-decade bug reports, scripts, and analyses on \href{https://github.com/HadiJahanshahi/WaybackMachine}{\textcolor{black}{GitHub}}.


\bibliographystyle{elsarticle-num-names} 
\bibliography{bib}

\end{document}